\documentclass[5p]{elsarticle}
\pdfoutput=1
\usepackage{graphicx,hyperref,natbib}
\usepackage[usenames,dvipsnames,svgnames,table]{xcolor}

\usepackage{amsmath}

\newcommand{\runz}{\textsc{runz}}
\newcommand{\autoz}{\textsc{autoz}}
\newcommand{\marz}{\textsc{Marz}}
\newcommand{\angstrom}{\mbox{\normalfont\AA}}

\begin{document}

\begin{frontmatter}

\title{\marz{}: Manual and Automatic Redshifting Software}

\author[affil1,affil2]{S.~R.~Hinton\corref{mycorrespondingauthor}}
\cortext[mycorrespondingauthor]{Corresponding author}
\ead{samuelreay@gmail.com}
\author[affil1]{Tamara M. Davis}
\author[affil3]{C. Lidman}
\author[affil4]{K. Glazebrook}
\author[affil5]{G. F. Lewis}

\address[affil1]{School of Mathematics and Physics, The University of Queensland, QLD 4072, Australia}
\address[affil2]{ARC Centre of Excellence for All-sky Astrophysics (CAASTRO)}
\address[affil3]{Australian Astronomical Observatory, North Ryde, NSW 2113, Australia}
\address[affil4]{Centre for Astrophysics and Supercomputing, Swinburne, University of Technology, Hawthorn, VIC 3122, Australia}
\address[affil5]{Sydney Institute for Astronomy, School of Physics, A28, The University of Sydney, NSW, 2006, Australia}

\begin{abstract}
The Australian Dark Energy Survey (OzDES) is a 100-night spectroscopic survey underway on the Anglo-Australian Telescope using the fibre-fed 2-degree-field (2dF) spectrograph.  We have developed a new redshifting application \marz{} with greater usability, flexibility, and the capacity to analyse a wider range of object types than the \runz{} software package previously used for redshifting spectra from 2dF. \marz{} is an open-source, client-based, Javascript web-application which provides an intuitive interface and powerful automatic matching capabilities on spectra generated from the AAOmega spectrograph to produce high quality spectroscopic redshift measurements. The software can be run interactively or via the command line, and is easily adaptable to other instruments and pipelines if conforming to the current FITS file standard is not possible. Behind the scenes, a modified version of the \autoz{} cross-correlation algorithm is used to match input spectra against a variety of stellar and galaxy templates, and automatic matching performance for OzDES spectra has increased from 54\% (\runz{}) to 91\% (\marz{}). Spectra not matched correctly by the automatic algorithm can be easily redshifted manually by cycling automatic results, manual template comparison, or marking spectral features.
\end{abstract}


\end{frontmatter}

\section{Introduction}

Redshift determination is a key component in many cosmological surveys. Whether the goal is to analyse supernovae, large scale structure, peculiar velocities, lensing, or a host of other interesting astronomical phenomenon, it is critical that the redshifts of target objects are determined to the highest resolution and free of unknown systematic effects. Of interest in this paper is the use of spectroscopic data to determine redshift, and prior spectroscopic surveys, such as the Two Degree Field \citep[2dF,][]{LewisCannonTaylor2002}, Six Degree Field  \citep[6dF,][]{JonesSaundersColless2004},  Wigglez \citep{Drinkwater2010}, Sloan Digital Sky Survey \citep[SDSS,][]{SmeeGunnUomoto2013}, Galaxy and mass Assembly  \citep[GAMA,][]{HopkinsDriverBrough2013} and Deep Extragalactic Evolutionary Probe 2  \citep[DEEP2,][]{NewmanCooper2013}, have used a variety of different software solutions and pipelines to attain redshift measurements.

These solutions have attacked the problem from multiple angles, with feature matching \citep{colless2001, kurtz1998rvsao, GarilliFuana2010}, $\chi^2$ minimisation \citep{aihara2011eighth, BoltonSchlegel2012, GlazebrookOfferDeeley1998, NewmanCooper2013}, and cross correlation \citep{baldry2014galaxy, colless2001, GarilliFuana2010, kurtz1998rvsao, StoughtonLupton2002, TonryDavis1979} the common three methodologies employed. Usually these solutions are bespoke software packages designed for a specific instrument.  We examine the redshifting requirements of two new surveys in this paper, OzDES and 2dFLenS, and from these requirements evaluate existing solutions and detail the creation of a new software package now in use by the surveys. The resulting \marz{} software is a web-application that, thanks to its ease of use (drag and drop a FITS file), may be adaptable as a generic viewer of spectroscopic fits files.\\

The Australian arm of the Dark Energy Survey (OzDES) is a five-year, 100-night spectroscopic survey using the Anglo-Australia Telescope (AAT), and aims to provide accurate spectroscopic redshifts of Type Ia supernova hosts, photo-$z$ targets, luminous red galaxies, emission line galaxies, and radio galaxies between redshift ranges of $0.1 \leq  z \leq 1.2$, in addition to spectroscopic measurements of active galaxy nuclei and quasars in the redshift range of $0.3 \leq z \leq 4.5$ \citep{fang2015}. The 2dFLenS survey\footnote{\url{http://astronomy.swin.edu.au/~cblake/2dflens_proposal.pdf}} aims to measure redshift-space distortions over 985 square degrees in its 50-night spectroscopic survey using with AAT.  The process of extracting redshifts from spectroscopic measurements is thus required to be robust across a wide variety of target types, and since our aim is primarily to acquire redshifts, our survey will be most efficient when we can extract redshifts from low signal-to-noise spectra.

Given the above motivation, we drew a set of minimum requirements a modern software replacement would need to satisfy:
\begin{enumerate}
\item The automatic matching has to be accurate and reliable, when compared to existing redshifting solutions.
\item The interface should be intuitive to allow fast manual checking and correction or verification by the user.
\item The software should be easy to install, operating system independent, and be able to be updated without user prompting.
\end{enumerate}
This paper will detail how the \marz{} software satisfies the above requirements. In Section~\ref{sec:prior} we briefly review prior software to provide motivation for a new redshifting application then in Sections~\ref{sec:platform} and \ref{sec:format} we justify our choice of software platform and input FITS file format respectively. Sections \ref{sec:algorithm} and \ref{sec:templates} detail the redshift matching algorithm and templates used.  The performance of our algorithm is assessed in Section~\ref{sec:perf}. In Section \ref{sec:interface} we explain how to utilise the software using both the interactive user interface and command line interface. Conclusions are presented in Section \ref{sec:conclusion}.

\section{Prior Software} \label{sec:prior}

A variety of redshifting solutions have been previously developed, with much of the developed software being spectrograph or survey specific. A large amount of prior redshifting software is descendent from the cross correlation algorithms implemented by \citet{TonryDavis1979}, in which they digitalised the analog techniques used by Griffin \cite{griffin1967photoelectric}. The software utilised previously by the OzDES team,  \runz{}, was originally written for use by the 2dF galaxy redshift survey \citep{colless2001}, and was modified by \citet{saunders2004} for use in the WiggleZ survey. It is primarily against this modified version of \runz{} that we compare results to in this paper.  Unfortunately, there are several reasons why neither the \runz{} software package, nor other available redshifting software packages, were sufficient for use in the OzDES survey. The foremost problem with using the existing software solution was the inclusion of more varied target object types at a lower signal-to-noise ratio than many prior surveys.\footnote{By design the survey maximises the number of of galaxies redshifted by taking the minimum quality spectrum required to acquire a redshift of each.}  The resulting decrease in automatic matching capacity by the legacy software created an undesirable workload for the members of the OzDES team to manually redshift observed spectra. Additionally, the legacy nature (fortran, pgplot, cshell, figaro libraries and starlink libraries) of the \runz{} code base makes code updates difficult, installation and usage complicated, especially for new users faced with having to compile, install and learn the software. The sum of these factors prompted a search for alternative software, leading to the development of a modern software replacement.

The main alternative considered was the popular RVSAO 2.0 software package, which utilises cross correlation techniques, in addition to providing feature matching capacity \citep{kurtz1998rvsao}. Unfortunately, the software operates only in command line and without any interactive user input. The 2010 release of the \verb+xcsao+ program (the cross correlation matching algorithm in RSVAO 2.0) takes up to 67 input parameters \citep{parameters2}, which, whilst allowing great specificity in redshifting, introduces technical overhead to cosmology groups attempting to use the software. Due to this, and the lack of modern interface, this software did not satisfy OzDES's requirements.

As of the eighth data release (DR8) of the Sloan Digital Sky Survey (SDSS), the redshifting software used in their survey utilises a minimum $\chi^2$ algorithm \citep{aihara2011eighth}. Prior to DR8, a cross correlation algorithm was used  \citep{sdss6}, where the same results were found when matching using the different methods for 98\% of spectra \cite{aihara2011eighth}, as expected, given that maximising the cross correlation strength gives a minimum $\chi^2$ value \citep{GlazebrookOfferDeeley1998}. This potential software system also does not satisfy the OzDES requirement of easy manual redshifting and confirmation. 

Finally, the GAMA survey utilises the \autoz{} code for redshifting low-redshift galaxy and stellar spectra \citep{baldry2014galaxy} by cross correlating input spectra with a range of templates. The \autoz{} IDL code does not feature a user interface, however the algorithm utilised outperformed the \runz{} algorithm for all spectra types barring quasar spectra. Unfortunately, IDL introduces an unwanted impediment to utilising the software, as it must be licensed. Due to the dependence on IDL and lack of an interface, the \autoz{} software did not meet the OzDES requirements. 

Given prior software was unable to satisfy all OzDES requirements, new software was developed.

\begin{figure*}[t]
	\centering
	\includegraphics[width=\textwidth]{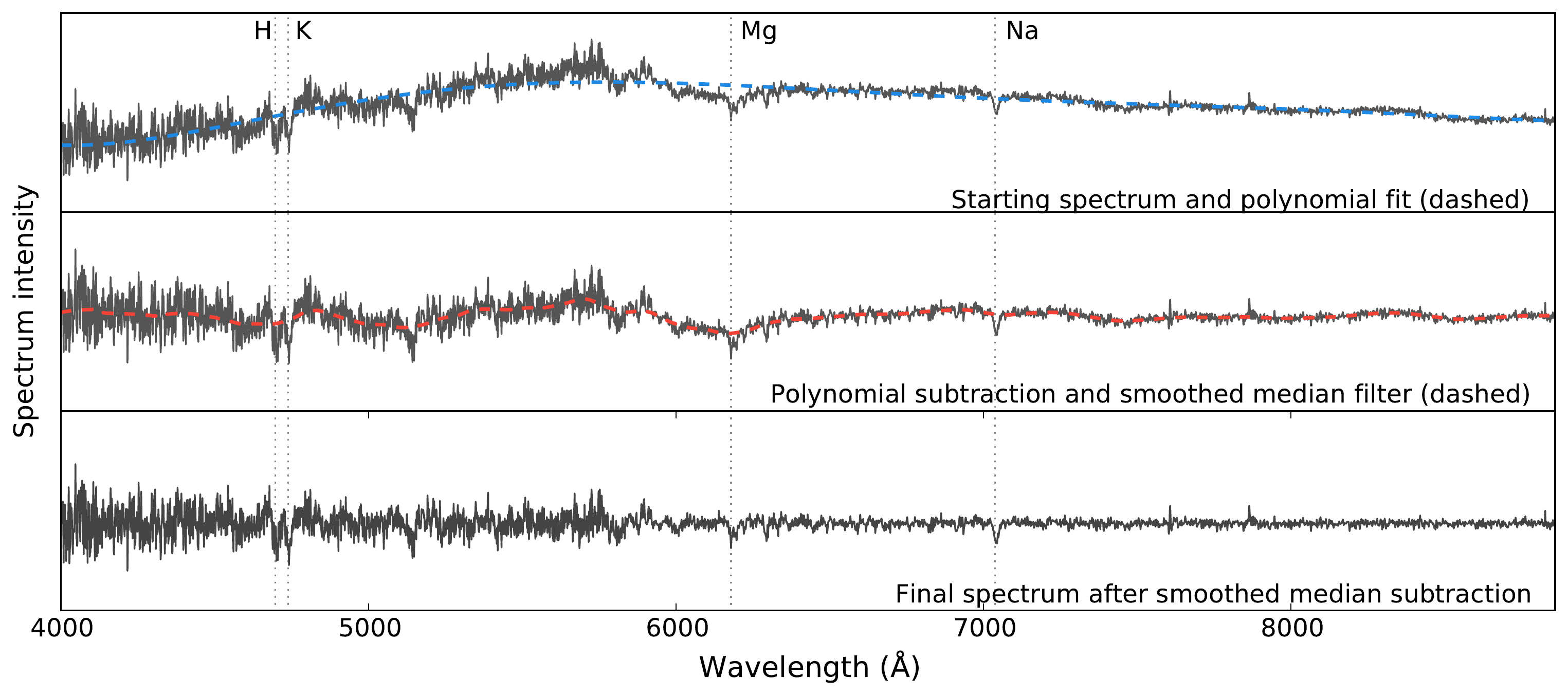}
	\caption{The top subfigure shows an example input spectrum from which the continuum will be removed. The sixth degree fitted polynomial is shown dashed in this subplot, and the spectrum after subtraction of this polynomial fit is shown in the middle subplot, where we can see that broad continuum features are removed. The middle subplot also shows the output of the smoothed median filtering (dashed), and spectrum after subtraction of this filter is shown in the bottom subplot, where we can see even fine continuum detail has been removed from the spectrum. We have deliberately chosen to show a spectrum that was processed with an earlier version of the 2dF data reduction pipeline (2dfdr), since spectra that were produced by this version of 2dfdr often have inaccurate continua with complex shapes. Later versions of 2dfdr more faithfully produce the continua of galaxies.}
	\label{fig:continuum}
\end{figure*}

\section{Platform} \label{sec:platform}

We chose to implement the matching software as a web application, since this allows access to the software from any laptop or desktop with an internet connection with no installation. The interface utilises Google's AngularJS  framework \citep{angularjs} for its application scaffold, as AngularJS allows dynamic two-way binding between interface and application variables for easy user interface creation, easy server communication, and has a vast amount of existing resources publicly available.

To supplement AngularJS and allow rapid prototyping, existing interface element libraries were imported to reduce the amount of reimplementation of common elements and boilerplate code required to produce a functioning application. To this end, Angular UI's Bootstrap \citep{angularUI} reimplementation of Bootstrap\citep{bootstrap} components was added to the project code base, to allow both the functionality of AngularJS and rapid prototyping of using Bootstraps pre-made components. As an example, this allowed Marz's Usage page, with its accordion layout, to be created in minutes, instead of the hours that would be required if no pre-existing components were available.

Data processing in browsers has only recently become possible due to HTML5's Web Worker API,\footnote{\url{https://html.spec.whatwg.org/multipage/workers.html}} which allows for multi-threaded processing by sending tasks to independent workers. Communication to Web Workers and any future potential server communication will use JSON (JavaScript Object Notation) format \citep{bray2014javascript2}, and conformation to the REST (Representational State Transfer) interface \citep{windley11rest} will allow for easy end-point construction by combining the instant serialization and deserialization of any Javascript object via JSON with the simple but structured API system provided by REST. This also has the benefit of making all messages human-readable and and able to be linked to web services easily due to existing support in all modern server frameworks for REST API's. The applications primary challenge to overcome with data processing was a lack of scientific Javascript libraries. This forced many basic mathematical and scientific functions to be reimplemented from scratch or translated from Python or IDL libraries, creating unwanted development overhead. Reimplementation and translation errors were checked via the creation of multiple test suites.

The local application state is preserved via utilisation of local storage, where local state is preserved in JSON format. This gives the benefit that results are not lost when exiting the application or refreshing the browser, negating one of the major disadvantages of stateful web applications. Changes to application are persisted via setting cookie properties instead of using local storage, to provide a concrete demarcation between user preferences and user redshifting data.

File saving functionality was added through the use of FileSaver.js \citep{save2}.

The code base, named \marz{}, is hosted publicly on GitHub,\footnote{\url{https://github.com/Samreay/Marz}} allowing for open issue management, feature requests, open collaboration, forking of the project and instant web-hosting.\footnote{\marz{} can be found at \url{http://samreay.github.io/Marz/}} As a web page, \marz{} updates automatically, and changes to the matching algorithm and output are reflected in an internal variable which stores the software version. Significant upgrades will also be released as product versions via the tagging capacity of git.

\begin{figure*}[t]
	\centering
	\includegraphics[width=\textwidth]{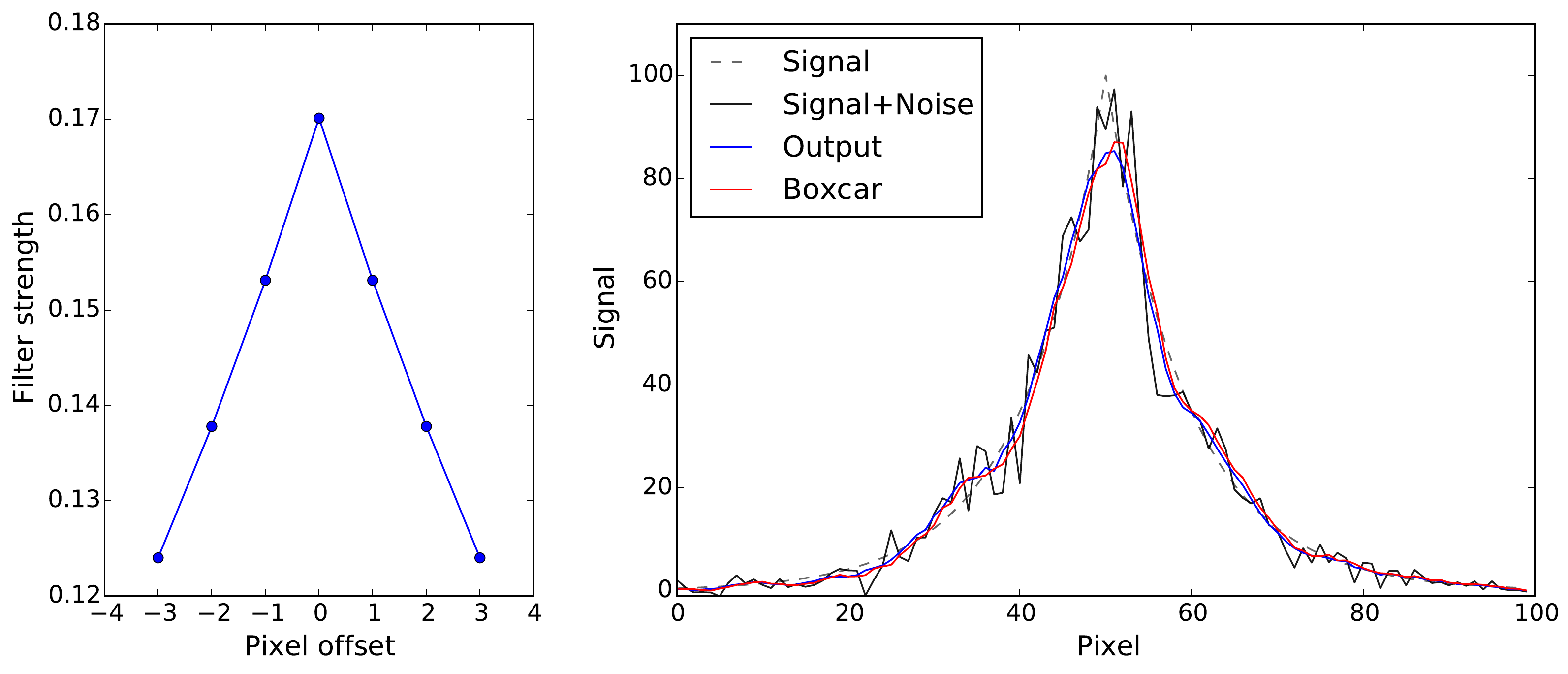}
	\caption{The left hand subplot shows the smoothing convolution filter used in \marz{} on the quasar spectra. On the right hand side, the black dashed line represents an underlying signal, modelled as an exponential decay. Independent Poisson noise is added to this signal to give the algorithm input (shown in black), and the output of the convolution of this signal with the smoothing filter is shown in blue. We can see that, despite the occurrence of subpeaks, the original signal peak is recovered, unlike the peak discovered when using boxcar smoothing (a uniform convolution), which is shown in red. This is due to the increased contribution at lower pixel separation, and increases the chance that a smoothed peak will be located at the same pixel in the noisy data.}
	\label{fig:rolling}
\end{figure*}

\begin{figure*}
	\centering
	\includegraphics[width=\textwidth]{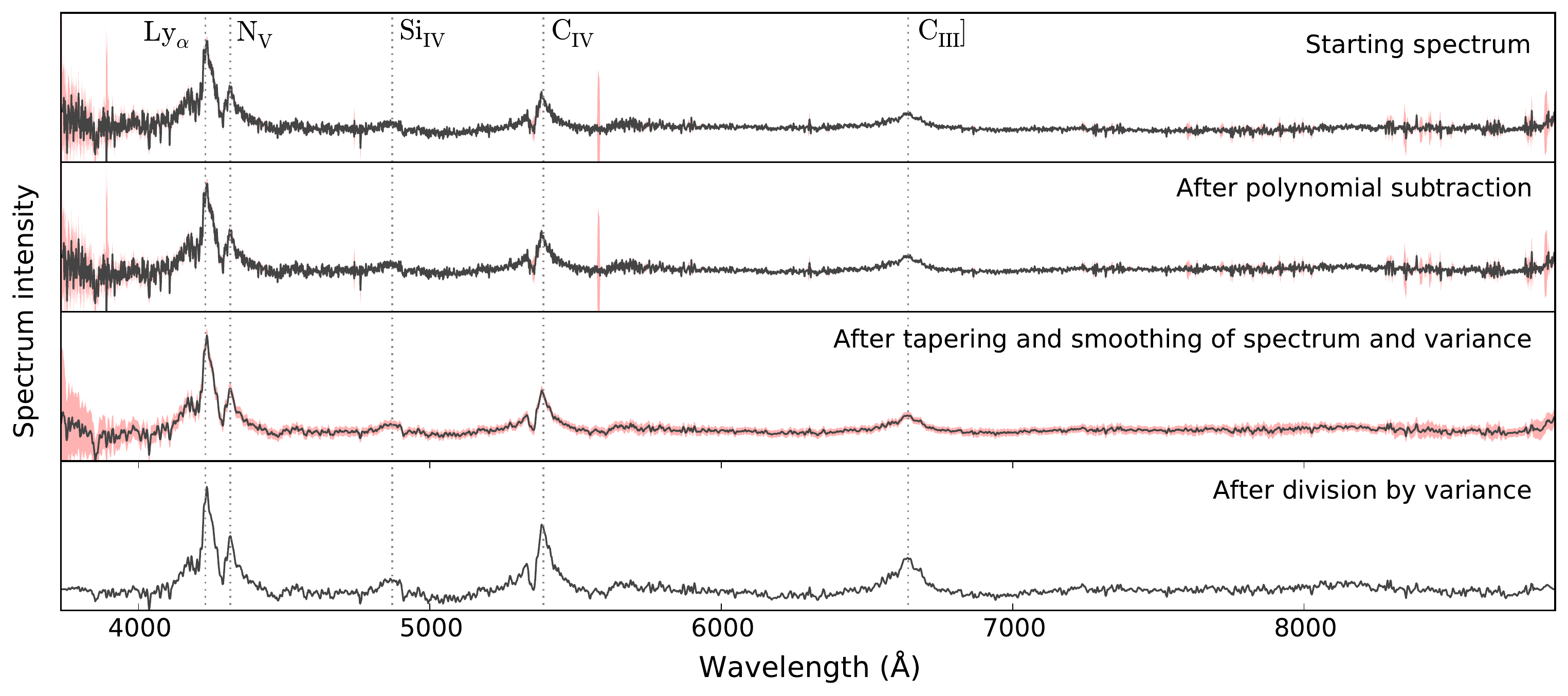}
	\caption{The quasar processing steps illustrated. The top panel depicts in input spectrum intensity, with variance shown in red. The polynomial subtraction removes the almost constant continuum and is shown in the second panel from the top. The third panel shows the output after several steps, including an initial tapering of the spectrum, rolling point mean of the spectrum. The variance undergoes median filtering, boxcar smoothing and addition of minimal variance as described in text. The final spectrum after dividing the intensity by the variance is shown in the bottom panel. We can see that the spectrum is cleaner, is properly tapered, and emission features are accentuated.}
	\label{fig:quasarProcess}
\end{figure*}

\section{FITS file format} \label{sec:format}

In this section, we detail the FITS file format that can be consumed by \marz{}. FITS files are loaded into the application via the \verb|fitsjs| library \citep{amit_kapadia_2015_16707}, and the data extraction algorithm then searches for extension names. Spectrum intensity is expected to found in either the primary header data unit (HDU) or one named \textit{intensity}. Spectrum variance is searched for using the extension name \textit{variance}, and similarly the \textit{sky} and \textit{fibres} should contain the sky spectrum and details on the fibres respectively. Although the primary use case of the software is with the AAOmega spectrograph (which uses 400 fibres at a time), the software can still be used for single spectra, either by omitting the fibres data or having only one row in the binary table. Whilst spectra can be viewed with only the intensity present, the matching algorithms require variance as well. The sky and fibre extensions are optional. The intensity and variance data should be present as images of $m\times n$ dimensions, where $m$ is the number of pixels in the spectrum and $n$ are the number of spectra in the file. The sky spectrum can take on a similar $m\times n$ format, or simply be present in an array of length $m$, which loads the same sky spectrum for all spectra.

If the \textit{fibres} extension exists, it should be as a binary table of $n$ rows. \marz{} searches for the columns TYPE, NAME, RA, DEC, MAGNITUDE, and COMMENT. Spectra marked with types other than `P' are removed from analysis (so as to not analyse fibres used for sky subtraction or calibration), and spectra marked with the comment `PARKED' are also removed. Priors can be enforced by specifying an object type in the comment field, and the effect of object type per template can be modified in the \verb;templates.js;, where the property \verb;weights; is used to map object types to numeric weights. These priors are used to increase or decrease the weight of specific templates for specific object types; as an example the type `AGN\_reverberation' is weighted to increase matching strength against the quasar template. If the types already defined in \marz{} are insufficient for a given survey, more can be added by raising an issue on the Github project, or by forking the project itself and directly editing the \verb;tempates.js; file. In Section \ref{sec:perf}, we compare the results of \marz{} and \runz{} for low signal-to-noise data (both using weights), and compare performance for high signal-to-noise without weights for either \marz{} or \runz{}.

\marz{} does not require each intensity array to have an explicitly declared counterpart wavelength array; the wavelength of the input spectra can be determined in multiple ways. \marz{} will search for an image extension named \textit{wavelength}, and if the extension exists, the image data will be loaded as wavelengths. The data can take an $m\times n$ format or simply be an array of length $m$ (similar to the sky data). If the extension is not found, \marz{} will check the primary header for cards with which to construct the wavelength array. The array is constructed such that a reference wavelength \verb;CRVAL1; (corresponding to pixel \verb;CRPIX1;) with linear pixel separation (either \verb;CDELT1; or \verb;CD1_1;) can be used to create a linear array of wavelength values. The wavelengths can be present in linear form (units of Angstroms), or if flag \verb;LOGSCALE; is set in the primary header, wavelength values are taken to be $\log_{10}(\lambda)$, such that a pixel value of $3$ would correspond to $1000\angstrom$. By default, \marz{} assumes the wavelengths presented are taken in air and will thus shift them into a vacuum reference frame. If this has already been done, \verb;VACUUM; can be set and \marz{} will not vacuum shift the input wavelengths.\\

Unlike many prior software implementations, \marz{} does not apply heliocentric velocity corrections by default. OzDES stack spectra from multiple runs in order to build up signal-to-noise, and therefore the spectra have to be put a consistent wavelength solution prior to stacking. This means the heliocentric correction (which differs from night to night) must be done before the spectra are stacked, and thus before \marz{} receives them. However, heliocentric corrections can be enabled in \marz{} by setting the header property \verb;DO_HELIO; to true. Upon finding this flag, \marz{} will correct for the heliocentric velocity, which requires the header to contain the modified Julian date of exposure (\verb;UTMJD;), epoch of exposure (\verb;EPOCH;), and the observatory's longitude, latitude and altitude (\verb;LONG_OBS;, \verb;LAT_OBS;, and \verb;ALT_OBS; respectively). The longitude and latitude are taken to be in degrees, and the altitude in meters above sea level. Furthermore, heliocentric velocity correction requires each spectrum have an associated right ascension and declination (\verb;RA; and \verb;DEC;), which would live in the \textit{fibres} extension as explained above. A correction into the rest frame of the CMB can also be performed, by setting the header \verb;DO_CMB; to true. Marz will then compute the CMB velocity using the \verb;RA; and \verb;DEC; values. The values for \verb;EPOCH; and \verb;RADECSYS; can also be specified - if they are not specified Marz defaults to 2000 and FK5 respectively. Given a computed heliocentric velocity $v_{hel}$ and peculiar velocity of the 3K background rest frame $v_{\rm CMB}$, the correct redshift $z$ is related to the observed redshift $z_{\rm obs}$ by
\begin{equation}
(1 + z) = \frac{(1 + z_{\rm obs})}{\left(1 - \frac{v_{\rm hel}}{c}\right)\left(1 - \frac{v_{\rm CMB}}{c}\right)}.
\end{equation}

Example FITS files can be downloaded on the \textit{Usage} section of the \marz{} application to provide file examples for the above specifications.

\begin{figure}[t]
	\centering
	\includegraphics[width=\columnwidth]{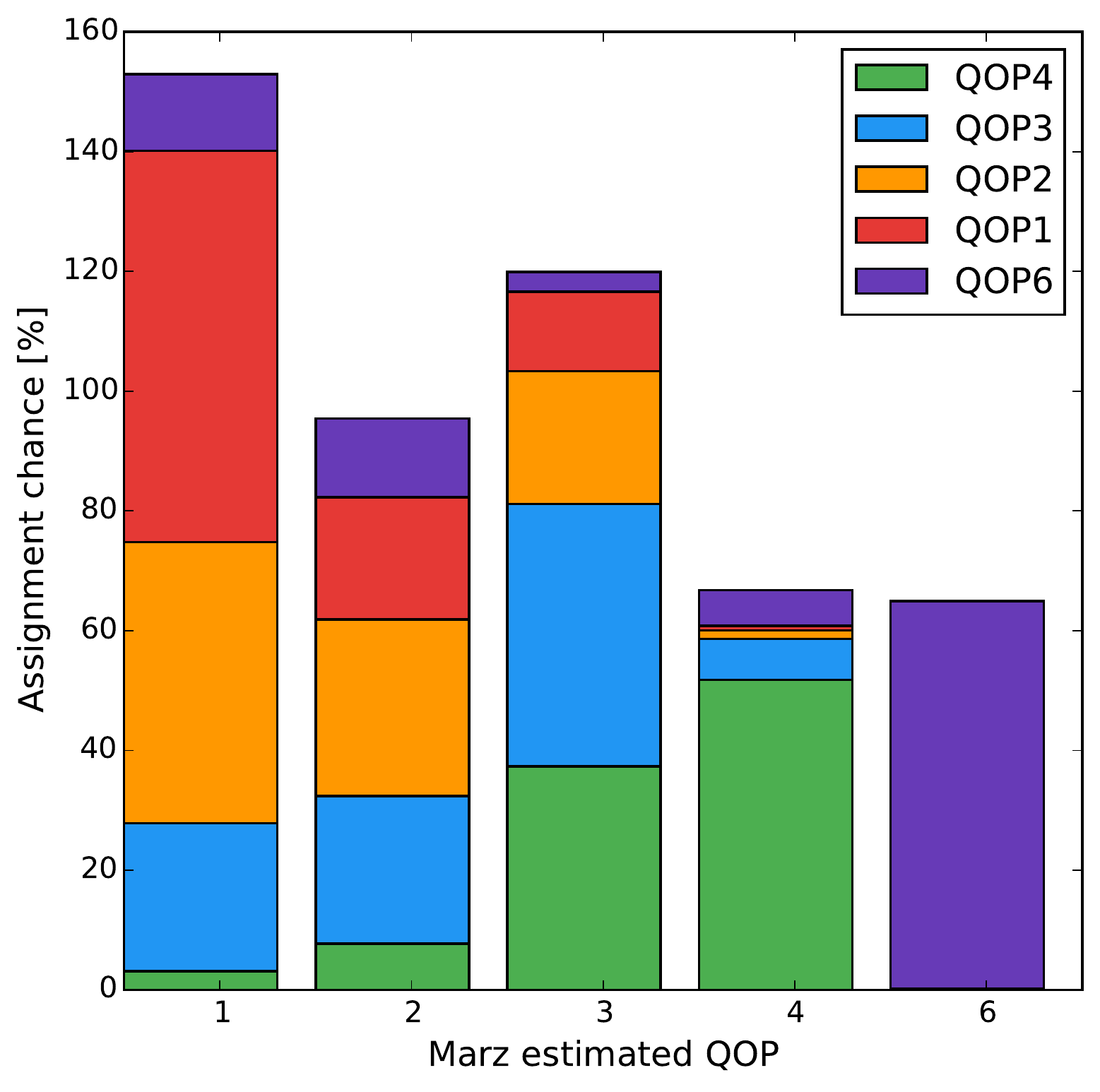}
	\caption{The probability of \marz{} assigning a suggested quality differentiated by the quality assigned by a human redshifter. Importantly, a QOP1 spectrum and a QOP2 spectrum respectively have a 0.8\%  and 1.4\% chance of being classified as a QOP4 spectra. However, these mismatches are primarily due to errors in the data reduction process which introduce false features. From the data analysed, 37\% of actual QOP4 spectra are classified as QOP3, where the automatic QOP algorithm has assigned a lower value due to the presence of a strong secondary peak in the cross correlation values, an occurrence illustrated in Figure \ref{fig:xcors}.}
	\label{fig:autoqop}
\end{figure}

\section{Matching Algorithm} \label{sec:algorithm}

The algorithm that takes an observed spectrum and measures the redshift is the heart of any redshifting program. The matching algorithms in \marz{} utilise a modified version of the \autoz{} algorithm implemented by \citet{baldry2014galaxy}. In light of the success of $\chi^2$ algorithms in modern surveys \citep{BoltonSchlegel2012} an initial $\chi^2$ algorithm was developed, but was consistently outperformed by the cross correlation algorithm and discarded. Performance of the $\chi^2$ algorithm could be increased by improving normalisation between the template and input spectra, but this was too computationally intensive when implemented. FITS file input from the AAOmega spectrograph undergo two distinct steps of processing: (1) the preprocessing stage to clean the data and (2) the matching process to align the observed spectra with template spectra, simultaneously finding the best-fit object type and shifting it to the best-fit redshift.

The preprocessing stage is designed to remove any bad pixels and cosmic rays from the data before being returned back to the interface, so that the user can manually redshift using the cleaned data.
\begin{itemize}
\item \textbf{Bad pixels} are defined when the intensity spectrum is NaN, negative or exceeds a certain configurable threshold, or if the variance spectrum for the pixel is negative.
\item \textbf{Cosmic rays} are identified via second neighbouring pixels exceeding thirty standard deviations from the mean, with lower thresholds than $30\sigma$ often mistaking strong emission lines as cosmic rays. 
\end{itemize}
Bad pixels have their intensity replaced with the mean to four pixels either side of the flagged pixel. Pixels flagged as a cosmic ray have a 9 pixel window (centered on the flagged pixel) replaced by a constant value, which is given by the mean intensity of pixels in a 19 pixel window (centered on the cosmic ray), discounting pixels in the cosmic ray. These pixel values were the minimum window found to produce sufficient quality means and remove the majority of cosmic rays.

The continuum is initially subtracted via the method of rejected polynomial subtraction, where a 6-degree polynomial is iteratively fitted to the spectrum and, as with \autoz{}, all points greater than 3.5 standard deviations from the mean are removed from the fitting process. As soon as an iteration discards no extra pixels, or after fifteen iterations (to ensure the final array of values is not excessively sparse), the loop is terminated and the final polynomial should closely follow the continuum, and is thus subtracted out.

This initial round of continuum subtraction is not intended to be high enough quality for the automatic matching process, it is done to give the user the option of manually redshifting spectra without continuum, allowing them to focus on the emission and absorption features of the spectrum without the broad shape of the continuum to distract. In order to limit the effect that singular emission features can have on spectrum matching, all features are clipped at a distance of 30 standard deviations from the mean.\\

\begin{figure*}[t]
	\centering
	\includegraphics[width=\textwidth]{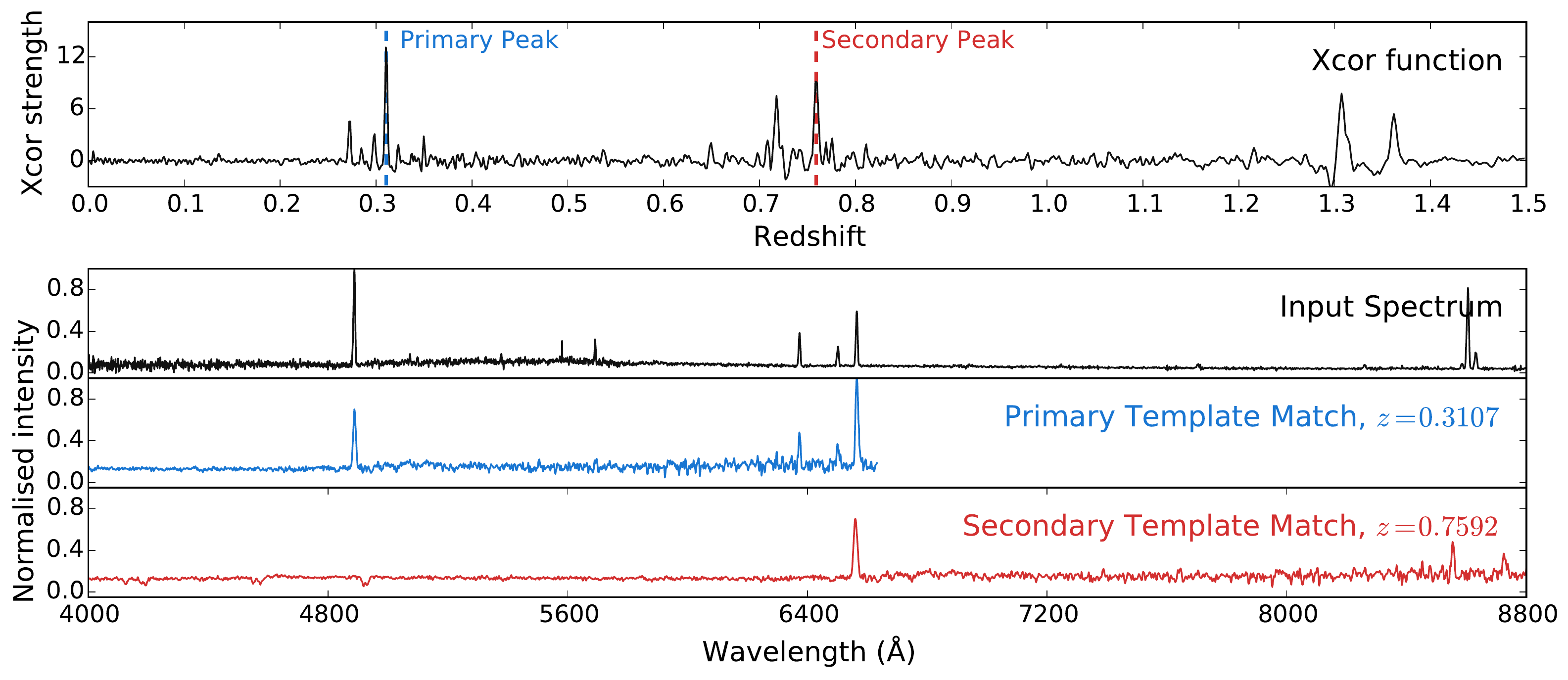}
	\caption{A high quality emission line galaxy spectrum matched against the High Redshift Star Forming Galaxy template. The two strongest cross correlation peaks and the corresponding redshifted template have been displayed beneath the original spectrum for illustrative purposes.}
	\label{fig:xcors}
\end{figure*}

The preprocessing algorithm detailed above takes an input intensity and variance spectrum, where the variance spectrum is typically computed in the data pipeline using Poisson and readout noise. The output of the preprocessing algorithm is an adjusted intensity and variance spectrum, and this becomes the input of the matching algorithm. In the matching algorithm, \marz{} first duplicates the intensity spectrum so that two copies exist internally. This is necessary because the matching of the broad-featured quasar template differs to matching of the other templates, and the copy of the intensity spectrum used for quasar matching shall now be referred to as the quasar spectrum, and the other spectrum - used to match all other templates, shall be referred to as the general spectrum.

\subsection{General spectra}

The general spectrum undergoes a cosine taper, and then undergoes a second step of continuum subtraction, where a boxcar smoothed median filter (121 pixel window of boxcar smoothing and 51 pixel window median filter) is subtracted from the spectrum, with pixel values following \autoz{} and correspond to a wavelength distance of 125 and 53 Angstroms respectively for typical AAOmega spectra. The cosine tapering of the spectrum ensures that features are not introduced at points of discontinuity at the ends of the spectrum in the case the input spectrum is zero-padded.  This second step of continuum subtraction is not applied to the quasar spectrum, as the fineness of the smoothed median subtraction would result in broad features being completely removed from the spectrum. Figure \ref{fig:continuum} illustrates each step in this process, in addition to the original polynomial continuum subtraction.

The general spectrum then has error adjustment applied, where each pixel has its variance set to the maximal value of itself and the variance of its two neighbouring pixels. Following \citet{baldry2014galaxy}, this is to allow for uncertainty in the sky subtraction and any underestimation of errors next to sky lines. The variance spectrum is then widened again, where each pixel is set to a maximum of its original value or 70\% of a local median filter, which serves to remove from the variance any points of sufficiently low variance that division of the intensity by the variance would create a fake emission feature.  The intensity of the general spectrum is divided by the variance spectrum to down-weight pixels with higher uncertainty. 

\subsection{Quasar spectra}

Departing from the \autoz{} algorithm, the quasar spectrum undergoes smoothing via a rolling-point exponential decay mean function of window width 7 pixels, with exponential decay factor of $0.9$, such that each point becomes
\begin{align}
x_i = \left( \sum_{k = -3}^3 0.9^{|k|}   \right)^{-1} \sum_{j = -3}^3 x_{i+j} 0.9^{|j|},
\end{align}
where both the size of the window and strength of the exponential decay are not sensitive to small changes. Pixel values were selected to maximise matching efficiency over test spectral data. This method of smoothing was selected and tuned so that the location of broad peaks remained unchanged whilst still providing sufficient smoothing of the spectrum to increase similarity to the template. This convolution is illustrated in Figure \ref{fig:rolling}.

As we wish to preserve broad features found in quasar spectra, we require the quasar error spectrum to be sufficiently smooth that broad features are not destroyed when we apply the variance onto the spectrum intensity. A median filter is applied to the variance, and then the result is smoothed with boxcar smoothing. The strength of these filters was not found to impact results, so long as fine pixel detail is removed from the variance plot. In order to preserve even more broad shape by creating a more uniform variance that does not have the possibility of creating false features by having small variance, the variance of the spectrum is increased by addition of five times the minimum spectrum variance, where again the results are not highly sensitive to the amount of variance added. The quasar intensity is then divided by the adjusted variance to produce a spectrum that retains broad features and shapes, but down-weights sections of higher variance which are commonly observed at wavelengths close the end of spectroscopic CCD range. The loss of resolution entailed by the variance modifications detailed above results in any sharp emission lines present in the spectrum to be more significant when compared to emission lines in the general spectrum, however the relative lack of sharp emission line features in the quasar template makes this issue not significant. These steps are illustrated in Figure \ref{fig:quasarProcess}.\\

\subsection{Quasar and general spectra}

Both the general and the quasar spectra undergo cosine tapering and root-mean-square normalisation, the former to remove ringing in a Fourier transform \citep[apodization;][]{kurtz1998rvsao}, with pixel width following \autoz{} but insensitive to change, and the latter to ensure comparable cross correlation values between different templates. The spectra are oversampled and then Fourier transformed. The quasar spectrum's transformation is then cross correlated with the quasar template, and all other templates are cross correlated with the general spectrum. Cross correlation results are then inverse transformed, with the inverse transformed array representing cross correlation strength over a redshift range. Peaks within allowed redshift ranges for each template are selected, and if prior information on the object type is accessible in the FITS file, the peaks for each template are then weighted. Peaks from all templates are then sorted together, and the ten highest correlation value peaks have a quadratic fit applied around the peak for sub-pixel determination of local maxima. The sub-pixel location of the peaks are converted into a redshift value, and these are returned to the user, with the highest peak representing the best automatic matching found. A potential quality is returned to the user, which is a function of the cross correlation strength of the two greatest peaks, $v_1$ and $v_2$ respectively. These peak values are used to construct a Figure of Merit (FOM), where we use the the functional form
\begin{align}
{\rm FOM} &= (v_1-\alpha)^{\beta} \times \frac{v_1}{v_2}. \label{eq:autoqop}
\end{align}
The values for $\alpha$, $\beta$ and the FOM cut offs shown in \eqref{eq:autoqop2} were informed off a likelihood distribution created to minimise the weighted number of misclassification. For example, a spectrum assigned a value of 4 for the quality operator (QOP), which represents 99.5\% confidence, was weighted higher than a misclassified QOP3 spectrum (95\% confidence), and so on for QOP2 spectra (signal present but unsure of type and/or redshift) and QOP1 (no confidence). We set parameters $\alpha$, $\beta$ and the FOM cutoff for each QOP as free variables in a Markov Chain Monte Carlo simulation, and then minimise the difference between AutoQOP and human assigned QOP. The approximate maximum likelihood of the marginalised distributions for this fit were selected as final values, such that $\alpha = 2.5$, $\beta = 0.75$ and the FOM boundaries are as given in \eqref{eq:autoqop2}.
\begin{align}
{\rm FOM} &= (v_1-2.5)^{0.75} \times \frac{v_1}{v_2} \label{eq:autoqop3}
\end{align}
\begin{align}
{\rm QOP} &= \begin{cases}6 & \text{if } {\rm FOM} > 4.5 \text{ and fit to a stellar template} \label{eq:autoqop2}\\
4 & \text{if } {\rm FOM} > 8.5  \\
3 & \text{if } {\rm FOM} > 4.5 \\
2 & \text{if } {\rm FOM} > 3 \\
1 & \text{otherwise} \end{cases}
\end{align}

This suggested QOP is not meant to replace human quality flags, but simply give the redshifter an estimate of spectrum quality before and during manual verification of the automatic result. The probability of agreement with a human redshifter is illustrated in Figure \ref{fig:autoqop}.  However, \marz{} can also be run in automatic mode, which outputs the AutoQOP without human intervention.

In this section, multiple numeric values have been presented, in terms of pixel width, polynomial degree and other limits. These values, chosen by testing them against AAOmega data, may not be optimal for other surveys which use other spectrographs and wavelength resolutions. To this point, all these values are contained in the \verb;config.js; file, and thus are easily modifiable upon forking the \marz{} project.

\section{Template Selection} \label{sec:templates}

\begin{figure*}[H]
	\centering
	\includegraphics[width=\textwidth]{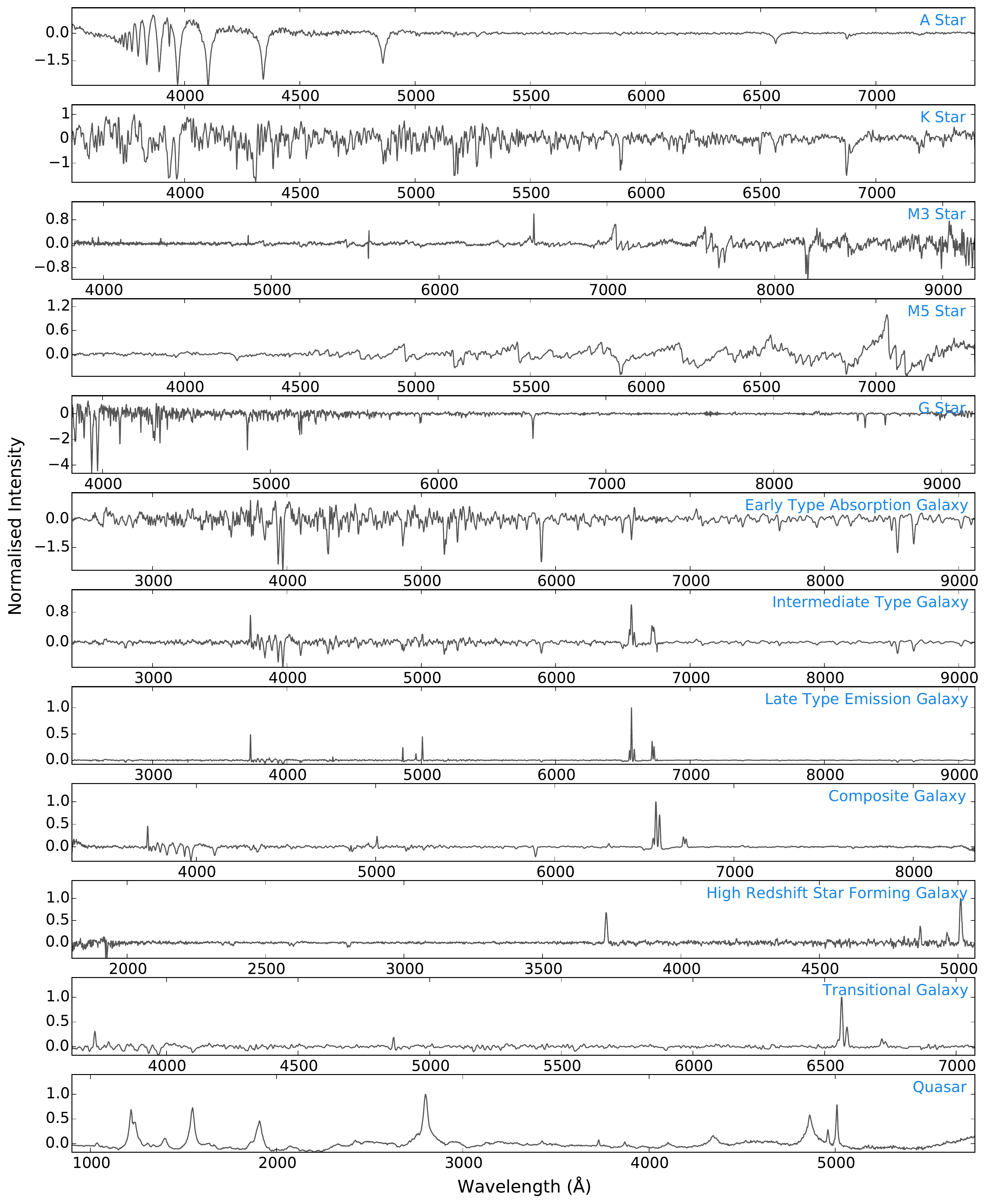}
	\caption{A visual display of the twelve templates currently in \marz{}, displayed after continuum subtraction.}
	\label{fig:templates}
\end{figure*}

It is common in automated matching systems to compare the input spectrum with a large number of templates. The majority of templates found in \marz{} were sourced from \runz{}, with original templates from 2dF \citep{colless2001}, WiggleZ \citep{Drinkwater2010} and the Gemini Deep Deep Survey \citet{abraham2004}. Other templates were sourced from \autoz{} \citep{baldry2014galaxy}, where the templates in \autoz{}  consist of galaxy eigenspectra from \citet{BoltonSchlegel2012} and stellar spectra from SDSS DR2 \citep{SubbaRao2002}. These templates were sorted, compared, and a selection of twelve representative templates were extracted, consisting of 5 stellar templates, 1 AGN template, and 6 galactic templates. Inclusion of a greater number of templates was found to have a minimal impact on matching performance - see Figure \ref{fig:high} for a comparison, and future improvements to matching performance will be looked for in the areas of eigentemplates or archetypal matching, rather than simply increasing the number of available templates.

Extra target types can be added in the future, however there are three challenges for \marz{} when attempting to handle a large number of templates. Firstly, users desired to be able to fully replicate the matching capacity of the automatic system when manually redshifting, and a large number of templates complicates the user interface and slows down the process of the user assigning an object type to the spectrum. It would be possible to only display to the user a restricted set of templates, however user feedback indicated this was an undesired solution. Due to the interpreted nature of Javascript and its lack of vector processing capability, computational performance is roughly an order of magnitude worse than on typical compiled code. As the computation time for each spectrum was roughly proportional to the number of templates to match, the number of templates was also kept relatively small to ensure that the automatic matching performance was still acceptable on low-end machines. The final concern when adding more templates is the download size of the web application, such that the size of the template dependency remains small enough to be easily redownloaded on page refresh. A potential solution to this is to enable javascript caching of the template file, such that it only needs to be downloaded once.

\clearpage

\section{Matching Performance} \label{sec:perf}

\begin{figure}[t]
	\centering
	\includegraphics[width=\columnwidth]{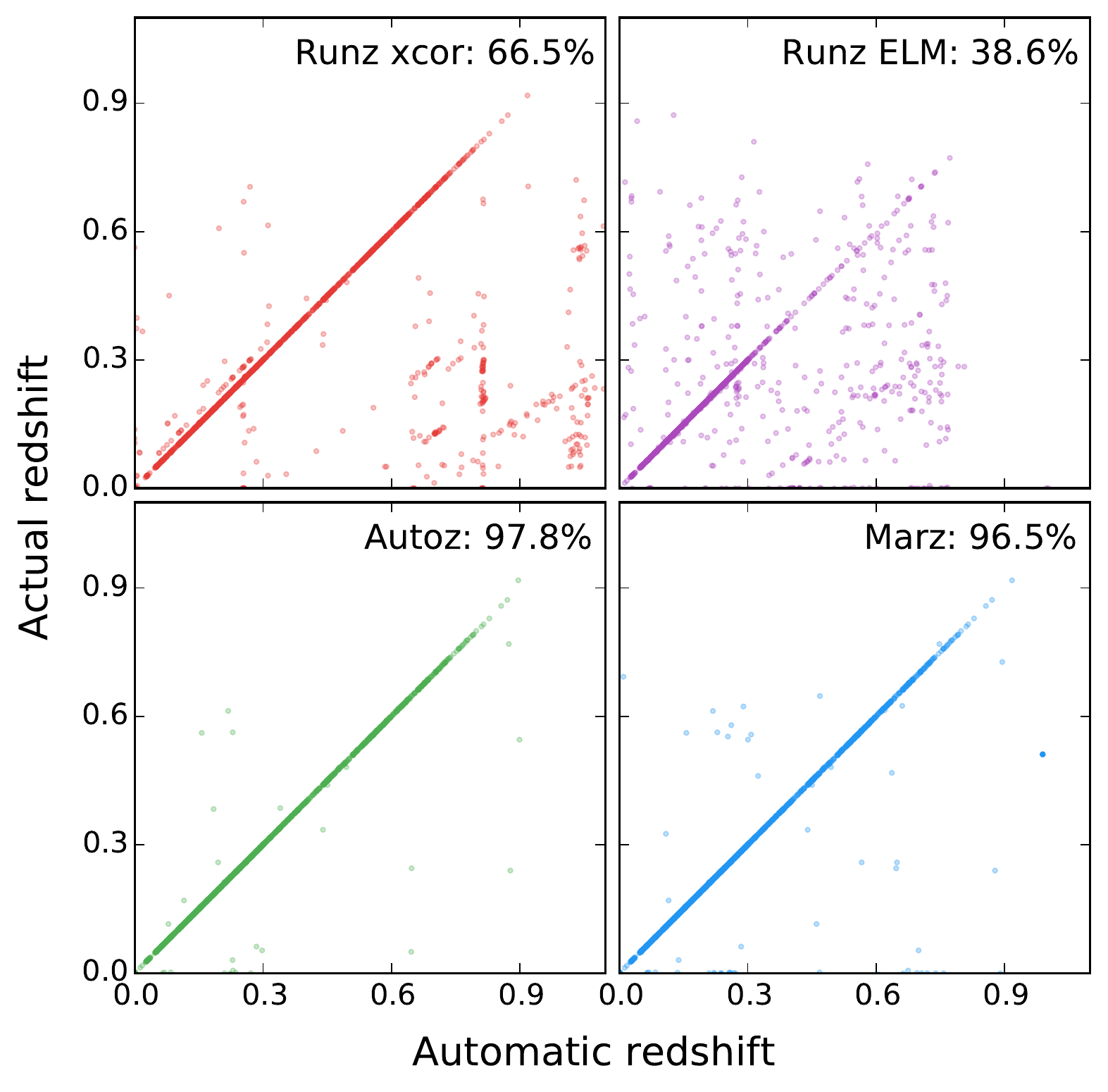}
	\caption{A comparison of matching efficiency using high signal-to-noise data from the 2dFLenS survey and a matching threshold of $\Delta \log(1+z) \leq 0.0015$, corresponding to a velocity difference of  $450$ km/s. 2217 QOP4 spectra from ten fields are compared in this plot. The vertical axes shows the redshift assigned by an experienced redshifter, and is taken to be correct in this comparison. The horizontal axes show the automatic results of the four algorithms being compared: the \runz{} cross correlation algorithm, the \runz{} emission line matching algorithm, \autoz{} and \marz{}. The success rate for each algorithm is shown in the top right of each subplot. The \marz{} algorithm and \autoz{} offer comparable accuracy for high quality spectra, with \autoz{} pulling ahead slightly due to an increased number of templates being used in the matching process. As discussed in section \ref{sec:templates}, compared to the \runz{} results, the inclusion of 52 templates in \autoz{}, in comparison to the 13 templates in \marz{}, did not provide a significant boost in redshift accuracy. The accuracy of all four algorithms displayed  could be further increased if template weighting were incorporated.}
	\label{fig:high}
\end{figure}

\begin{figure}[t]
	\centering
	\includegraphics[width=\columnwidth]{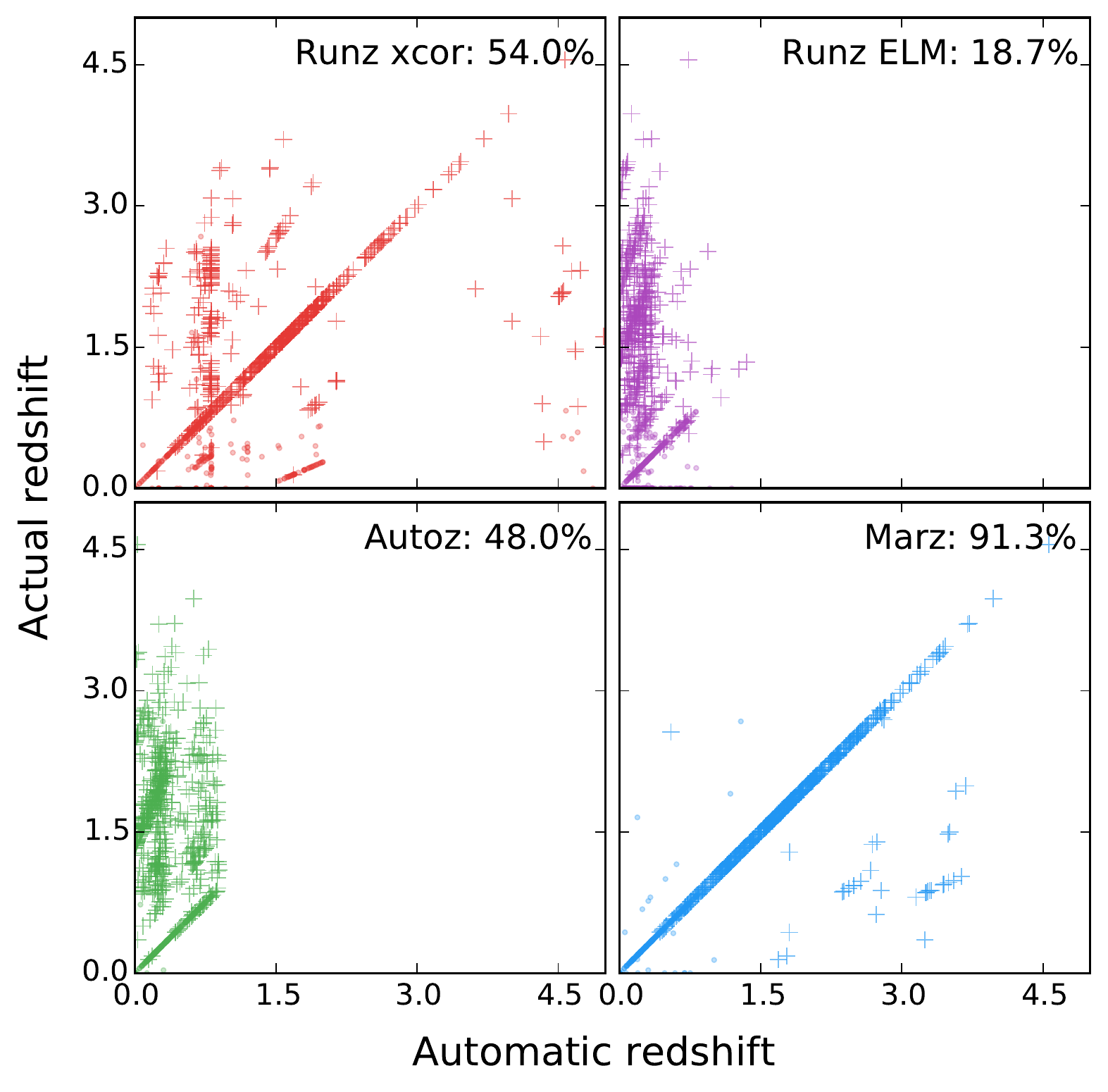}
	\caption{Low signal-to-noise, high-redshift data from the OzDES survey is used in this comparison of matching capability, where two spectra agree if their recession velocity is within 450 km/s of each other for standard galactic templates, or within 900 km/s of each other for AGN spectra. Respectively this corresponds to $\Delta \log(1+z) < 0.0015$ and $\Delta \log(1+z) < 0.003$. AGN matches are shown with the $+$ symbol, with other objects shown with a dot. Overall success rates are shown in the top right hand corner of each subplot. 1083 spectra from eight fields with a redshift range of up to 4.55 are used in this comparison. \runz{} emission line matching performs the worst, with strong diagonal lines of non-unity slope showing repeated spectrum feature misidentification. Vertical banding in the \runz{} cross correlation algorithm significantly impacts its effectiveness, and \autoz{}'s lack of high redshift templates and hard-coded $z=0.9$ cutoff make the comparison almost inapplicable. We should also note in this comparison that \runz{} and \marz{} were both run with template weighting enabled, giving them greater information than available to the \autoz{} algorithm. Given both these concerns, this plot should not be taken to show the algorithmic limits of the \autoz{} algorithm, it simply shows the success rate when run, without extra development or modification, against OzDES data. The success rate for the \autoz{} algorithm would improve if the algorithm were modified to remove the hard limit and template weighting implemented. Here the quasar specific matching algorithm used by \marz{} stands out, giving a success rate of over 90\% for the OzDES spectra.}
	\label{fig:low}
\end{figure}

Performance testing for \marz{} was conducted by looking at two sets of distinct data - one from the OzDES team with low signal-to-noise data at high redshift, and one from the 2dFLenS team with medium signal-to-noise data. In both cases, manual redshifting was performed by experienced redshifters in \runz{}, and the automatic matches produced by \runz{} and the automatic results returned by \marz{} were compared to the manually assigned redshift for all results assigned a QOP of 4. Comparisons were also made with the \autoz{} program, which is the software being used for the GAMA survey. These surveys have a smaller number of object types and smaller redshift ranges than OzDES, and therefore have simpler requirements for the redshifting software. These high and low signal-to-noise results are shown in Figures \ref{fig:high} and \ref{fig:low}. The redshifting accuracy of \marz{} for high signal-to-noise data gave the correct redshift for 97.4\% of QOP4 spectra, a failure rate far less than that offered by \runz{} and comparable to \autoz{}. For the low signal-to-noise (high redshift) OzDES data, the accuracy of \marz{} was $91.3$\%. This is in comparison to the best \runz{} algorithm giving a total accuracy of 54.6\%. The lower success rate of \autoz, 48.0\%, is because the redshift ranges and object types found in the OzDES data are outside the matching capacity of \autoz{}.

\begin{figure}[H]
	\centering
	\includegraphics[width=\columnwidth]{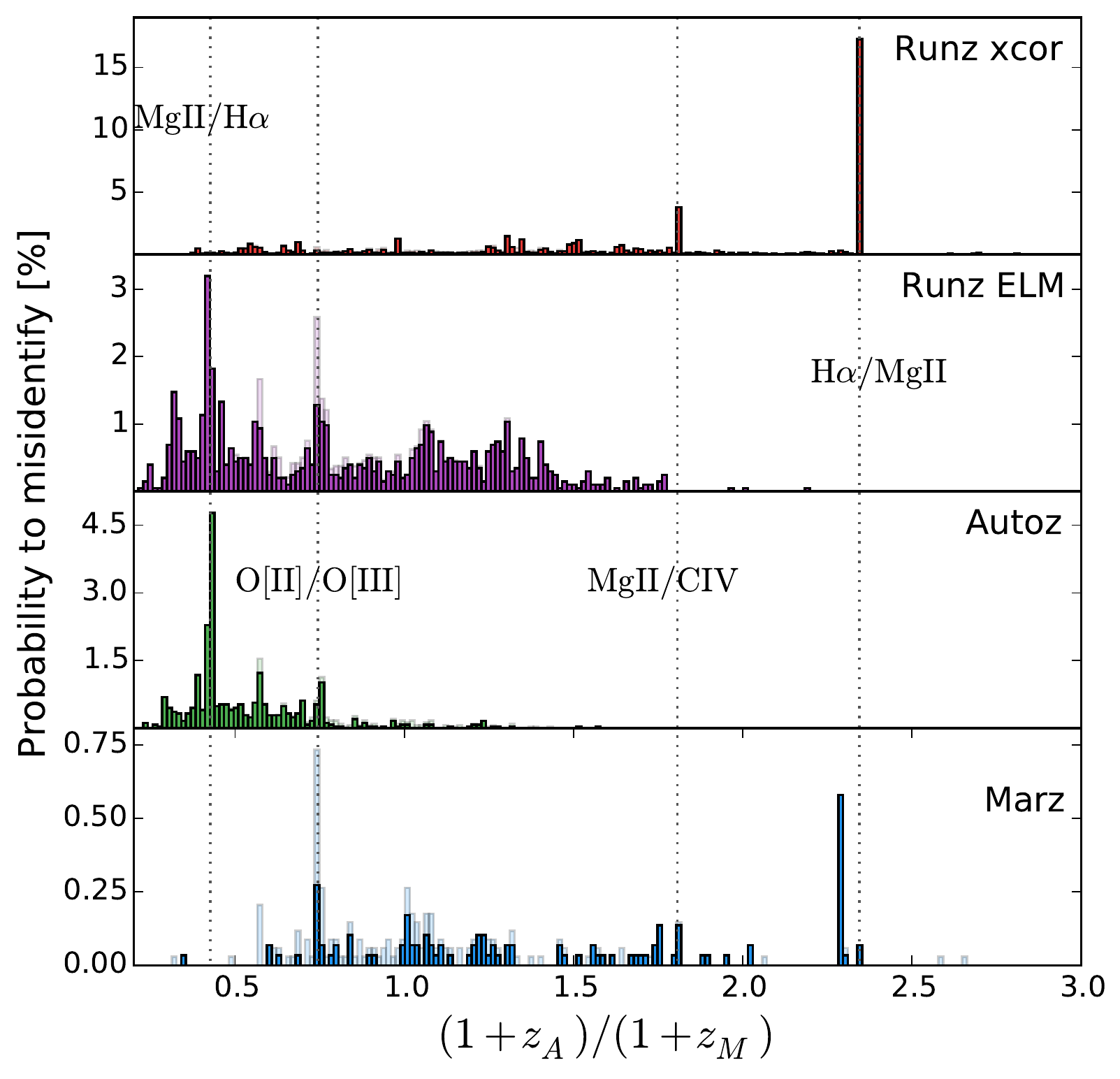}
	\caption{The percentage chance, per successfully assigned redshift of quality 3 or 4, of assigning an automatic redshift $z_A$ categorised as a catastrophic failure with respect to correct manual redshift $z_M$ (note that the manual redshift may be incorrect, and QOP4 spectra are those rated to be correct 99\% percent of the time). QOP4 misclassifications are shown in solid colour, and total QOP4+QOP3 misclassifications are shown in the translucent bars. A catastrophic failure, here defined as $\Delta \log(1 + z) > 0.0075$, corresponds to a velocity difference of $2\,250$ km/s and generally indicates misclassified features in the spectra. For an analysis of non-catastrophic failures, see Figure \ref{fig:systematic}. Peaks in the probability distribution generally represent misidentified spectral lines, and common misidentification ratios have had the corresponding spectral lines labelled, such that the first label, MgII/H$\alpha$ represents the MgII feature misidentified as H$\alpha$ instead. Performance for \runz{} cross correlation, \runz{} emission line matching, \autoz{} and \marz{} are shown in their own panel. Panel probability axes are not to scale with one another, and the area covered represents the total failure rate. The 2dFLenS and OzDES data from Figures \ref{fig:high} and \ref{fig:low} are combined in this analysis to give the greatest number of data points. The catastrophic failure rates for QOP4 spectra are given as follows: \runz{} cross correlation: 37.4\%, \runz{} emission line matching: 33.1\% failure rate, \autoz{}: 19.4\% failure rate, \marz{}: 3.77\% failure rate.}
	\label{fig:f4}
\end{figure}

In addition to the QOP4 successful recovery rates, we can check for catastrophic matching failures ($\Delta \log(1+z) \geq 0.0075$) by creating a normalised probability distribution describing the likelihood for spectral line misclassification. This has been done for QOP4 and QOP3 results, in Figure \ref{fig:f4}. Commonly misidentified spectral lines are labeled in the figure, and can be seen to be a source of misclassification across all algorithms. The common [O\textsc{ii}]/[O\textsc{iii}] misclassification is present in the \marz{} failure rates, especially for the QOP3+QOP4 results, however it accounts for only approximately only one in four hundred misclassifications when comparing with QOP4 only data, an thus does not represent a significant issue with the matching algorithm. Commonly misclassified features are also visible on Figure \ref{fig:high} and \ref{fig:low} as linear relationships off the diagonal.

\begin{figure}[t]
	\centering
	\includegraphics[width=\columnwidth]{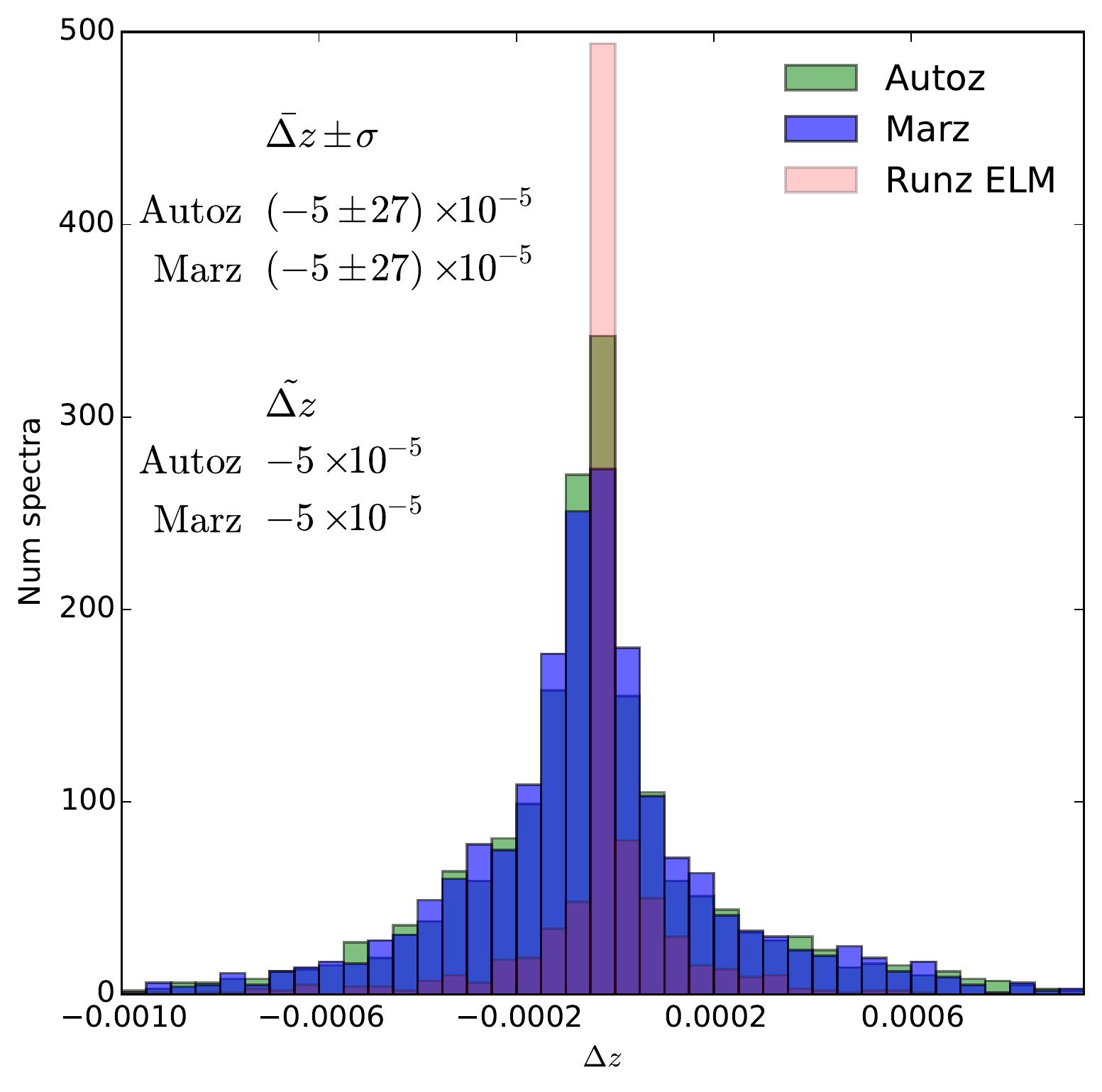}
	\caption{Systematic redshift offsets were investigated by examining the difference between manual redshifting (using the \runz{} cross correlation algorithm as the base) and the automatic results of redshifting algorithms. QOP4 spectra from OzDES and 2dFLenS were combined to give over three thousand secure manual redshifts, with the redshift distribution shown particulary for \marz{} and \autoz{}. The mean offset and median for both algorithms is to the order of $10^{-5}$ or less, with the redshift variance on order $10^{-4}$. The redshift mean and median values are small enough to be indicative that no systematic offset in redshift would be generated by utilising either algorithm.}
	\label{fig:systematic}
\end{figure}

We checked \marz{} for any systematic redshift offset by comparing the redshift result obtained in \marz{} against results obtained by using different, well tested software - \runz{} and \autoz{}. The redshift distribution found is shown in Figure \ref{fig:systematic}. The mean and median redshift offset from \marz{} ($4\times10^{-5}$ and $5\times10^{-6}$ respectively) are comparable to the offsets found with \autoz{}, and do not show any sign of a systematic offset. The variance in redshift results found in \marz{} ($\approx 3 \times 10^{-4}$) is also comparable to \autoz{}, and within the expected variance for AGN and galaxy targets (determined by repeated observation of the same target) of $\Delta z/(1+z) = 0.0015$ and $\Delta z/(1+z) = 0.0005$ respectively for the OzDES survey \citep{fang2015}. Due to the redshift variance between observations dominating any uncertainty in an individual match, \marz{} does not assign uncertainty to redshift results. Instead, uncertainty will be determined based on object type and the results of repeated observations, which is detailed in \citet[Table 4]{fang2015}.\\

\clearpage

\section{Interface} \label{sec:interface}

\subsection{Interactive Interface}

\begin{figure*}[h]
\centering
\includegraphics[width=\textwidth]{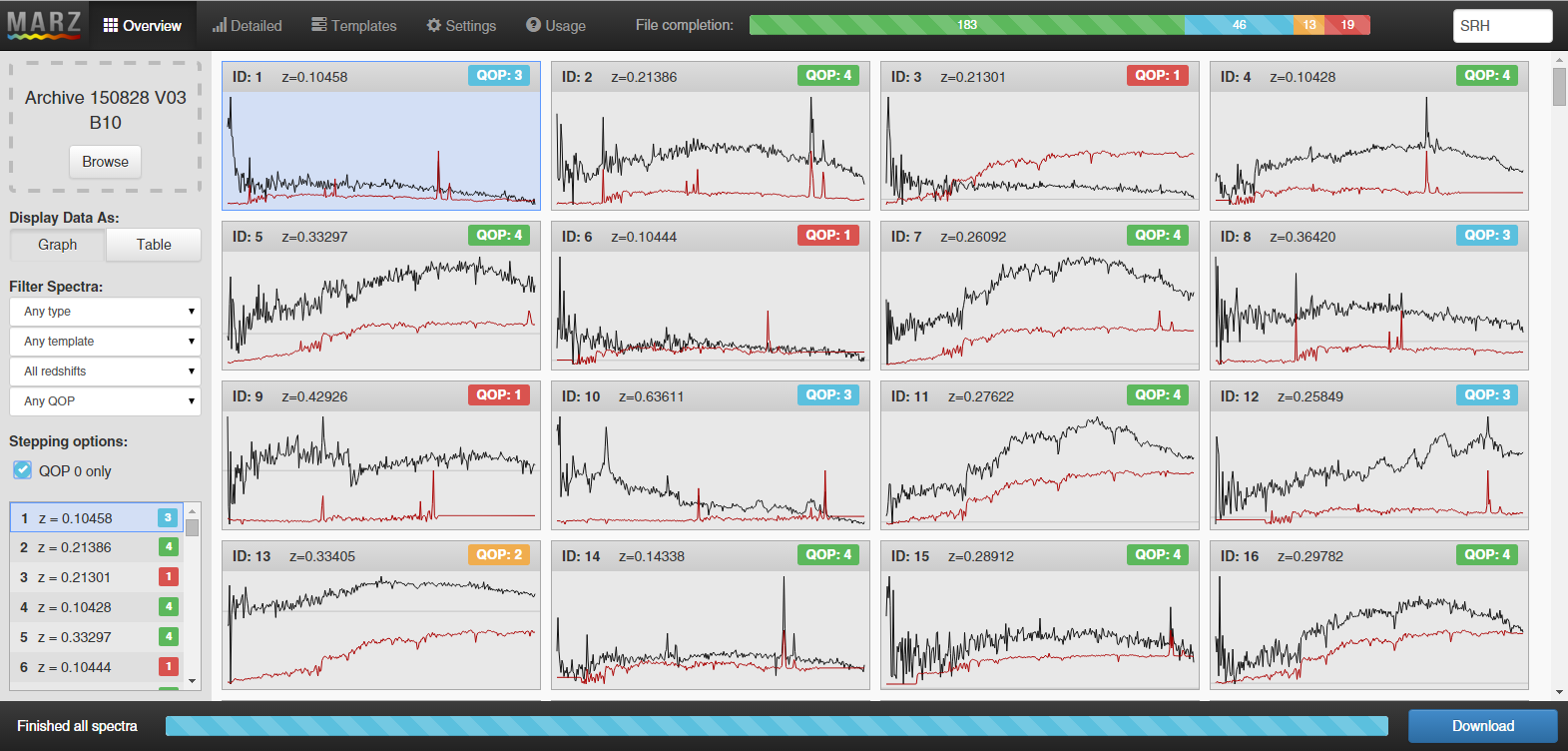}
\caption{The overview screen, showing data from a FITS file courtesy of Chris Blake and the 2dFLenS survey. Users can switch between a sortable tabular view and a graphic tile view, filter on object types, redshift ranges, templates and QOP values. The top of the screen shows the navigation menu, file completion progress bar and input for user initials. Visible at the bottom of the screen is the application footer, which shows the program's progress through automatic matching (automatically matched templates are shown in red in the graphical tiles). The bar changes colour depending on progress - green for preprocessing, red for matching and blue for completed. During the first two stages, a pause button is available to the user. If any results exist, a download button is available, which saves the current results to the file system.}
\label{fig:overview}
\end{figure*}

\begin{figure*}[h]
\centering
\includegraphics[width=\textwidth]{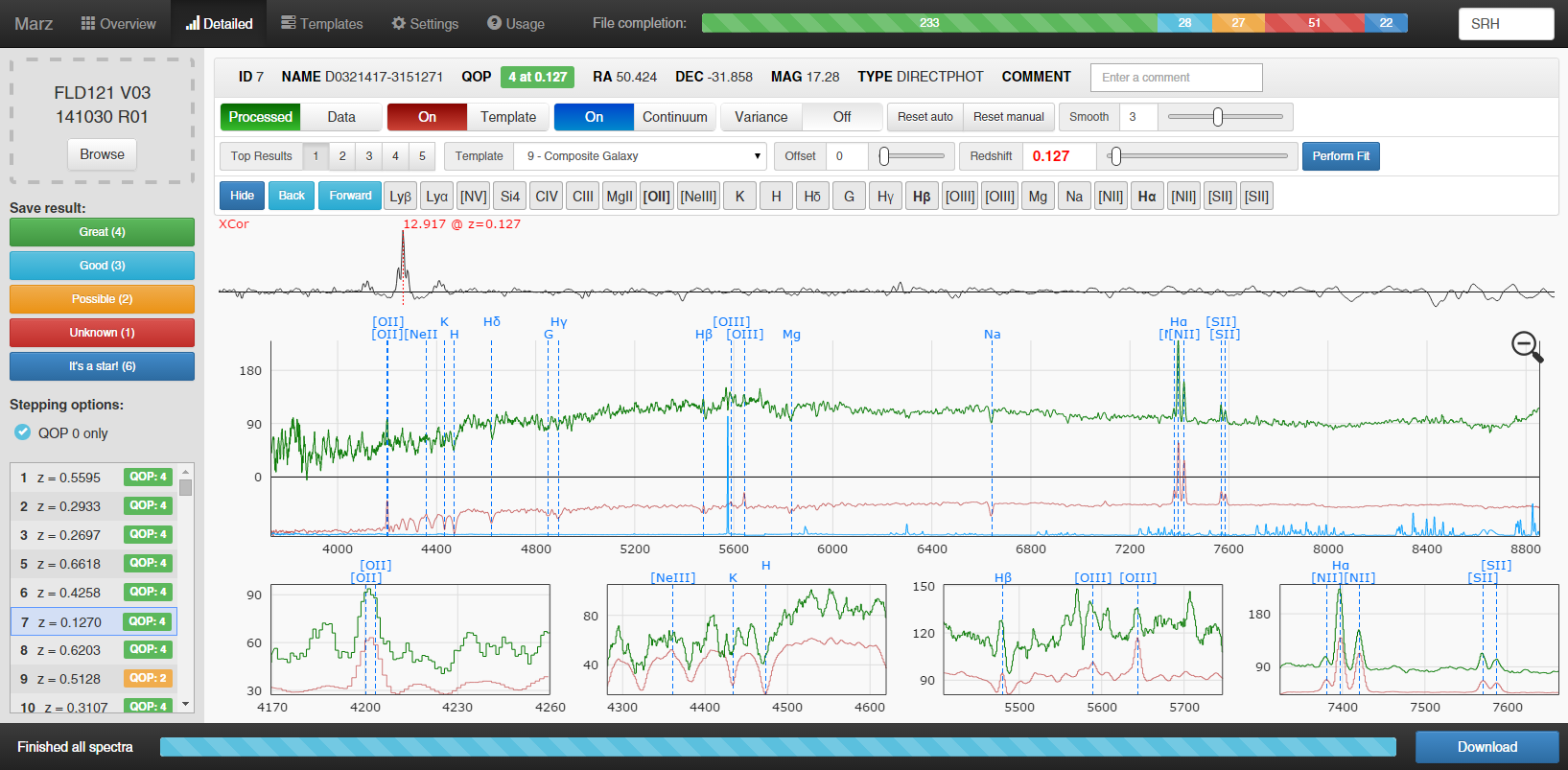}
\caption{The detailed matching screen, showing spectrum 7 seen in the Overview screen in Figure \ref{fig:overview}. The menus at the top of the page allow the user to toggle data on or off (variance, templates and whether to use the raw data or processed data). The menu bar also allows the user to reset to automatic or manual results, smooth the data, select which template to compare against, toggle between the best five automatic results, change the visual offset of the template and manually set the displayed redshift. The user can mark spectral lines by selecting a feature in the plot (either in the main plot or in any callout window) and then select the desired transition (either via keyboard shortcut or by selecting an option in the bottom row of the menu). Users can also change redshift by clicking on peaks in the cross correlation graph found between the spectra plot and the menu bars. Quality values for redshifts can also be assigned via keyboard shortcuts or via the vertical button menu on the left, and assigning a quality saves the result in the background and moves to the next spectrum. In the case where the ``QOP 0 only" option is selected in the left hand bar, the user is taken to the next spectrum without a quality flag set, or else it simply takes them to the next spectrum ordered by ID.}
\label{fig:detailed}
\end{figure*}

The interactive interface consists of five primary screens: the overview, detailed, templates, settings, and usage screen. The first two screens - the overview and detailed screens, are where users will spend the vast majority of their time, and thus screenshots of them have been provided in Figures \ref{fig:overview} and \ref{fig:detailed}. The overview screen provides users with a high level view of the spectra in the loaded FITS file, detailing what they have been matched to and the quality assigned to the matches. Filtering for this screen allows users to sort results or to filter by categories, for example only displaying matches of quality (QOP) 4 or all matches to quasar templates. Matching results can be downloaded as comma-separated variable (CSV) files, and those same files can be used to load the matching results back into \marz{} on different machines by simply dropping the results file into the program the same way as a user would load in a FITS file. This was added to allow easy verification of redshift results by different users on different machines. A progress bar at the top of the screen keeps track of current file completion and file quality.

The detailed screen allows for better verification of automatic matches and also offers the possibility of manually redshifting spectra. Verification of the on screen displayed redshift is done simply by assigning a QOP value, and the top five automatic matches can be cycled if the best match is visibly incorrect. Keyboard shortcuts are available for almost all actions, where key mappings are based off the shortcuts available in \runz{} in order to make transitioning from \runz{} to the \marz{} as easy as possible. Users can click on features in the detailed plot and then mark them as spectral lines. Matches can be updated by by automatically fitting to the best cross correlation peak value within a small deviation window. The user can also toggle whether to display the raw data or the preprocessed data, whether to render a template under the data plot, and whether to display continuum or not. Boxcar smoothing is available to help spectrum legibility.

The templates screen is mostly non-interactive, and simply displays all the templates used by the system with the option to enable or disable specific templates at will. The settings screen gives options to explicitly set how many processing threads to create, whether results should be saved in the background, and offers the ability to clear all saved results in the system, or to simply clear results for the currently loaded FITS. The usage page gives instructions on how to use the program, an explanation of the purpose of each screen, how to raise issues or feature requests via GitHub, and provides several example FITS files for users who simply want to test out the system without having to source a FITS file themselves. It also provides a list of keyboard shortcuts for those users whom are not familiar with \runz{}.

Two main error safeguards have been implemented in the program to stop unnecessary loss of work. The first is a confirmation request when attempting to close down the application, which solves the issue of closing the whole browser with an open tab of \marz{}. The second and more robust solution is to use the local storage capacity available in modern browsers to save results in the background after every automatic or manual redshift is assigned. This allows users to close the program, and resume where they left off simply by dragging the original FITS file back into the application.\\

\subsection{Command Line Interface} \label{sec:commandline}

A command line interface has also been developed for \marz{}. It requires  node.js\footnote{\url{https://nodejs.org/}}to be installed, and the two node dependencies are installed when running \verb;install.sh; - minimist\footnote{\url{https://github.com/substack/minimist}} and q\footnote{\url{https://github.com/kriskowal/q}}. The run script provided (\verb;marz.sh;) take a minimum single argument - the path of the file or folder to analyse. If the path points to a FITS file, Marz runs against the file. If the path specifies a folder, Marz will run against all FITS files inside the folder (not recursively). Specific configuration options, such as the output folder for \verb;.mz; files, can be configured either through the command line or by modifying \verb;autoConfig.js;. More detailed usage instructions and command line parameters are detailed in the project Github readme, or by the application itself when run without input parameters.\footnote{\url{https://github.com/samreay/Marz}}\\

\section{Conclusion} \label{sec:conclusion}
For both high and low signal-to-noise data, \marz{} performs consistently well at automatically redshifting spectra. \marz{} also provides an enhanced and intuitive user experience, allowing the user to visualise spectra, and manually redshift spectra by cycling automatic matches, manually entering desired redshift, or marking emission or absorption features. The web-based nature of the application means that installation and updating are now no longer of any concern at all, and unlike many web-based applications, \marz{} also saves users' work in the background, so that work is not lost with unintentional browser closure.

The development of such a large and technical application, without significant supporting scientific libraries, has been a massive challenge. Whilst, from a usability and interface perspective, web-applications may excel, recent development of the ability to perform large computations on background processes via web workers means that the mathematical library support for Javascript is still in its infancy. Whilst this can be overcome with effort, it provides a significant barrier for application development. This barrier has been overcome with \marz{}, and as such \marz{} presents a strong redshifting application, and a large step forward in the demonstration of web frameworks as a platform for non-intensive computational analysis.

As \marz{} simply requires generic FITS files to work, it can be used for simple FITS file inspection and visualisation, or as a redshifting tool for other surveys.

\section*{Acknowledgements}
We would like to thank the OzDES and 2dFLenS teams for their feedback, input and user testing, especially the 2dFLenS lead, Chris Blake. The \marz{} template catalogue comes from the WiggleZ and SDSS template samples, and multiple external libraries have been utilised in the creation of this application: Google's AngularJS, Twitter's Bootstrap, AngularUI, Amit Kapadia's fitsjs, Corban Brook's Digital Signal Processing package, Eli Grey's FileSaver.js, Tom Alexander's regression.js package, NASA's astronomical IDL libraries, lodash and jQuery. Parts of this research were conducted by the Australian Research Council Centre of Excellence for All-sky Astrophysics (CAASTRO), through project number CE110001020.

\newpage 
\section*{References}
\bibliography{bibliography}

\begin{thebibliography}{29}
\expandafter\ifx\csname natexlab\endcsname\relax\def\natexlab#1{#1}\fi

\bibitem[{{Abraham} {et~al.}(2004){Abraham}, {Glazebrook}, {McCarthy},
  {Crampton}, {Murowinski}, {J{\o}rgensen}, {Roth}, {Hook}, {Savaglio}, {Chen},
  {Marzke}, \& {Carlberg}}]{abraham2004}
{Abraham}, R.~G. {et~al.} 2004, \aj, 127, 2455

\bibitem[{Adelman-McCarthy {et~al.}(2008)Adelman-McCarthy, Ag{\"u}eros, Allam,
  Prieto, Anderson, Anderson, Annis, Bahcall, Bailer-Jones, Baldry,
  {et~al.}}]{sdss6}
Adelman-McCarthy, J.~K. {et~al.} 2008, The Astrophysical Journal Supplement
  Series, 175, 297

\bibitem[{Aihara {et~al.}(2011)Aihara, Prieto, An, Anderson, Aubourg, Balbinot,
  Beers, Berlind, Bickerton, Bizyaev, {et~al.}}]{aihara2011eighth}
Aihara, H. {et~al.} 2011, The Astrophysical Journal Supplement Series, 193, 29

\bibitem[{{Angular UI}(2015)}]{angularUI}
{Angular UI}. 2015, Angular UI Bootstrap,
  \url{https://github.com/angular-ui/bootstrap}

\bibitem[{Baldry {et~al.}(2014)Baldry, Alpaslan, Bauer, Bland-Hawthorn, Brough,
  Cluver, Croom, Davies, Driver, Gunawardhana, {et~al.}}]{baldry2014galaxy}
Baldry, I. {et~al.} 2014, Monthly Notices of the Royal Astronomical Society,
  441, 2440

\bibitem[{{Bolton} {et~al.}(2012){Bolton}, {Schlegel}, {Aubourg}, {Bailey},
  {Bhardwaj}, {Brownstein}, {Burles}, {Chen}, {Dawson}, {Eisenstein}, {Gunn},
  {Knapp}, {Loomis}, {Lupton}, {Maraston}, {Muna}, {Myers}, {Olmstead},
  {Padmanabhan}, {P{\^a}ris}, {Percival}, {Petitjean}, {Rockosi}, {Ross},
  {Schneider}, {Shu}, {Strauss}, {Thomas}, {Tremonti}, {Wake}, {Weaver}, \&
  {Wood-Vasey}}]{BoltonSchlegel2012}
{Bolton}, A.~S. {et~al.} 2012, \aj, 144, 144, 1207.7326

\bibitem[{{Bootstrap}(2014)}]{bootstrap}
{Bootstrap}. 2014, Bootstrap, \url{http://getbootstrap.com/}, accessed:
  2014-08-23

\bibitem[{Bray(2014)}]{bray2014javascript2}
Bray, T. 2014, The JavaScript Object Notation (JSON) Data Interchange Format,
  \url{http://www.json.org/}

\bibitem[{{Colless} {et~al.}(2001){Colless}, {Dalton}, {Maddox}, {Sutherland},
  {Norberg}, {Cole}, {Bland-Hawthorn}, {Bridges}, {Cannon}, {Collins}, {Couch},
  {Cross}, {Deeley}, {De Propris}, {Driver}, {Efstathiou}, {Ellis}, {Frenk},
  {Glazebrook}, {Jackson}, {Lahav}, {Lewis}, {Lumsden}, {Madgwick}, {Peacock},
  {Peterson}, {Price}, {Seaborne}, \& {Taylor}}]{colless2001}
{Colless}, M. {et~al.} 2001, \mnras, 328, 1039

\bibitem[{Drinkwater {et~al.}(2010)Drinkwater, Jurek, Blake, Woods, Pimbblet,
  Glazebrook, Sharp, Pracy, Brough, Colless, Couch, Croom, Davis, Forbes,
  Forster, Gilbank, Gladders, Jelliffe, Jones, Li, Madore, Martin, Poole,
  Small, Wisnioski, Wyder, \& Yee}]{Drinkwater2010}
Drinkwater, M.~J. {et~al.} 2010, Monthly Notices of the Royal Astronomical
  Society, 401, 1429, arXiv:0911.4246

\bibitem[{{Garilli} {et~al.}(2010){Garilli}, {Fumana}, {Franzetti}, {Paioro},
  {Scodeggio}, {Le F{\`e}vre}, {Paltani}, \& {Scaramella}}]{GarilliFuana2010}
{Garilli}, B., {Fumana}, M., {Franzetti}, P., {Paioro}, L., {Scodeggio}, M.,
  {Le F{\`e}vre}, O., {Paltani}, S., \& {Scaramella}, R. 2010, \pasp, 122, 827,
  1005.2825

\bibitem[{{Glazebrook} {et~al.}(1998){Glazebrook}, {Offer}, \&
  {Deeley}}]{GlazebrookOfferDeeley1998}
{Glazebrook}, K., {Offer}, A.~R., \& {Deeley}, K. 1998, \apj, 492, 98,
  astro-ph/9707140

\bibitem[{{Google}(2014)}]{angularjs}
{Google}. 2014, Angularjs, \url{https://angularjs.org/}, accessed: 2014-08-23

\bibitem[{Grey(2014)}]{save2}
Grey, E. 2014, FileSaver.js, \url{https://github.com/eligrey/FileSaver.js/},
  accessed: 2014-09-09

\bibitem[{Griffin(1967)}]{griffin1967photoelectric}
Griffin, R. 1967, The Astrophysical Journal, 148, 465

\bibitem[{{Hopkins} {et~al.}(2013){Hopkins}, {Driver}, {Brough}, {Owers},
  {Bauer}, {Gunawardhana}, {Cluver}, {Colless}, {Foster}, {Lara-L{\'o}pez},
  {Roseboom}, {Sharp}, {Steele}, {Thomas}, {Baldry}, {Brown}, {Liske},
  {Norberg}, {Robotham}, {Bamford}, {Bland-Hawthorn}, {Drinkwater}, {Loveday},
  {Meyer}, {Peacock}, {Tuffs}, {Agius}, {Alpaslan}, {Andrae}, {Cameron},
  {Cole}, {Ching}, {Christodoulou}, {Conselice}, {Croom}, {Cross}, {De
  Propris}, {Delhaize}, {Dunne}, {Eales}, {Ellis}, {Frenk}, {Graham},
  {Grootes}, {H{\"a}u{\ss}ler}, {Heymans}, {Hill}, {Hoyle}, {Hudson}, {Jarvis},
  {Johansson}, {Jones}, {van Kampen}, {Kelvin}, {Kuijken},
  {L{\'o}pez-S{\'a}nchez}, {Maddox}, {Madore}, {Maraston}, {McNaught-Roberts},
  {Nichol}, {Oliver}, {Parkinson}, {Penny}, {Phillipps}, {Pimbblet}, {Ponman},
  {Popescu}, {Prescott}, {Proctor}, {Sadler}, {Sansom}, {Seibert},
  {Staveley-Smith}, {Sutherland}, {Taylor}, {Van Waerbeke}, {V{\'a}zquez-Mata},
  {Warren}, {Wijesinghe}, {Wild}, \& {Wilkins}}]{HopkinsDriverBrough2013}
{Hopkins}, A.~M. {et~al.} 2013, \mnras, 430, 2047, 1301.7127

\bibitem[{{Jones} {et~al.}(2004){Jones}, {Saunders}, {Colless}, {Read},
  {Parker}, {Watson}, {Campbell}, {Burkey}, {Mauch}, {Moore}, {Hartley},
  {Cass}, {James}, {Russell}, {Fiegert}, {Dawe}, {Huchra}, {Jarrett}, {Lahav},
  {Lucey}, {Mamon}, {Proust}, {Sadler}, \&
  {Wakamatsu}}]{JonesSaundersColless2004}
{Jones}, D.~H. {et~al.} 2004, \mnras, 355, 747, astro-ph/0403501

\bibitem[{Kapadia \& Merg1255(2015)}]{amit_kapadia_2015_16707}
Kapadia, A., \& Merg1255. 2015, fitsjs: 0.6.6

\bibitem[{Kurtz \& Mink(1998)}]{kurtz1998rvsao}
Kurtz, M.~J., \& Mink, D.~J. 1998, Publications of the Astronomical Society of
  the Pacific, 110, 934

\bibitem[{{Lewis} {et~al.}(2002){Lewis}, {Cannon}, {Taylor}, {Glazebrook},
  {Bailey}, {Baldry}, {Barton}, {Bridges}, {Dalton}, {Farrell}, {Gray},
  {Lankshear}, {McCowage}, {Parry}, {Sharples}, {Shortridge}, {Smith},
  {Stevenson}, {Straede}, {Waller}, {Whittard}, {Wilcox}, \&
  {Willis}}]{LewisCannonTaylor2002}
{Lewis}, I.~J. {et~al.} 2002, \mnras, 333, 279, astro-ph/0202175

\bibitem[{Mink(2010)}]{parameters2}
Mink, D. 2010, rvsao.xcsao Parameters,
  \url{http://tdc-www.harvard.edu/iraf/rvsao/xcsao/xcsao.par.html}, url date:
  2010-05-24

\bibitem[{{Newman} {et~al.}(2013){Newman}, {Cooper}, {Davis}, {Faber}, {Coil},
  {Guhathakurta}, {Koo}, {Phillips}, {Conroy}, {Dutton}, {Finkbeiner}, {Gerke},
  {Rosario}, {Weiner}, {Willmer}, {Yan}, {Harker}, {Kassin}, {Konidaris},
  {Lai}, {Madgwick}, {Noeske}, {Wirth}, {Connolly}, {Kaiser}, {Kirby},
  {Lemaux}, {Lin}, {Lotz}, {Luppino}, {Marinoni}, {Matthews}, {Metevier}, \&
  {Schiavon}}]{NewmanCooper2013}
{Newman}, J.~A. {et~al.} 2013, \apjs, 208, 5, 1203.3192

\bibitem[{{Saunders} {et~al.}(2004){Saunders}, {Cannon}, \&
  {Sutherland}}]{saunders2004}
{Saunders}, W., {Cannon}, R., \& {Sutherland}, W. 2004, Anglo-Australian
  Observatory Epping Newsletter, 106, 16

\bibitem[{{Smee} {et~al.}(2013){Smee}, {Gunn}, {Uomoto}, {Roe}, {Schlegel},
  {Rockosi}, {Carr}, {Leger}, {Dawson}, {Olmstead}, {Brinkmann}, {Owen},
  {Barkhouser}, {Honscheid}, {Harding}, {Long}, {Lupton}, {Loomis}, {Anderson},
  {Annis}, {Bernardi}, {Bhardwaj}, {Bizyaev}, {Bolton}, {Brewington}, {Briggs},
  {Burles}, {Burns}, {Castander}, {Connolly}, {Davenport}, {Ebelke}, {Epps},
  {Feldman}, {Friedman}, {Frieman}, {Heckman}, {Hull}, {Knapp}, {Lawrence},
  {Loveday}, {Mannery}, {Malanushenko}, {Malanushenko}, {Merrelli}, {Muna},
  {Newman}, {Nichol}, {Oravetz}, {Pan}, {Pope}, {Ricketts}, {Shelden},
  {Sandford}, {Siegmund}, {Simmons}, {Smith}, {Snedden}, {Schneider},
  {SubbaRao}, {Tremonti}, {Waddell}, \& {York}}]{SmeeGunnUomoto2013}
{Smee}, S.~A. {et~al.} 2013, \aj, 146, 32, 1208.2233

\bibitem[{{Stoughton} {et~al.}(2002){Stoughton}, {Lupton}, {Bernardi},
  {Blanton}, {Burles}, {Castander}, {Connolly}, {Eisenstein}, {Frieman},
  {Hennessy}, {Hindsley}, {Ivezi{\'c}}, {Kent}, {Kunszt}, {Lee}, {Meiksin},
  {Munn}, {Newberg}, {Nichol}, {Nicinski}, {Pier}, {Richards}, {Richmond},
  {Schlegel}, {Smith}, {Strauss}, {SubbaRao}, {Szalay}, {Thakar}, {Tucker},
  {Vanden Berk}, {Yanny}, {Adelman}, {Anderson}, {Anderson}, {Annis},
  {Bahcall}, {Bakken}, {Bartelmann}, {Bastian}, {Bauer}, {Berman},
  {B{\"o}hringer}, {Boroski}, {Bracker}, {Briegel}, {Briggs}, {Brinkmann},
  {Brunner}, {Carey}, {Carr}, {Chen}, {Christian}, {Colestock}, {Crocker},
  {Csabai}, {Czarapata}, {Dalcanton}, {Davidsen}, {Davis}, {Dehnen},
  {Dodelson}, {Doi}, {Dombeck}, {Donahue}, {Ellman}, {Elms}, {Evans}, {Eyer},
  {Fan}, {Federwitz}, {Friedman}, {Fukugita}, {Gal}, {Gillespie}, {Glazebrook},
  {Gray}, {Grebel}, {Greenawalt}, {Greene}, {Gunn}, {de Haas}, {Haiman},
  {Haldeman}, {Hall}, {Hamabe}, {Hansen}, {Harris}, {Harris}, {Harvanek},
  {Hawley}, {Hayes}, {Heckman}, {Helmi}, {Henden}, {Hogan}, {Hogg}, {Holmgren},
  {Holtzman}, {Huang}, {Hull}, {Ichikawa}, {Ichikawa}, {Johnston}, {Kauffmann},
  {Kim}, {Kimball}, {Kinney}, {Klaene}, {Kleinman}, {Klypin}, {Knapp},
  {Korienek}, {Krolik}, {Kron}, {Krzesi{\'n}ski}, {Lamb}, {Leger},
  {Limmongkol}, {Lindenmeyer}, {Long}, {Loomis}, {Loveday}, {MacKinnon},
  {Mannery}, {Mantsch}, {Margon}, {McGehee}, {McKay}, {McLean}, {Menou},
  {Merelli}, {Mo}, {Monet}, {Nakamura}, {Narayanan}, {Nash}, {Neilsen},
  {Newman}, {Nitta}, {Odenkirchen}, {Okada}, {Okamura}, {Ostriker}, {Owen},
  {Pauls}, {Peoples}, {Peterson}, {Petravick}, {Pope}, {Pordes}, {Postman},
  {Prosapio}, {Quinn}, {Rechenmacher}, {Rivetta}, {Rix}, {Rockosi}, {Rosner},
  {Ruthmansdorfer}, {Sandford}, {Schneider}, {Scranton}, {Sekiguchi}, {Sergey},
  {Sheth}, {Shimasaku}, {Smee}, {Snedden}, {Stebbins}, {Stubbs}, {Szapudi},
  {Szkody}, {Szokoly}, {Tabachnik}, {Tsvetanov}, {Uomoto}, {Vogeley}, {Voges},
  {Waddell}, {Walterbos}, {Wang}, {Watanabe}, {Weinberg}, {White}, {White},
  {Wilhite}, {Wolfe}, {Yasuda}, {York}, {Zehavi}, \&
  {Zheng}}]{StoughtonLupton2002}
{Stoughton}, C. {et~al.} 2002, \aj, 123, 485

\bibitem[{SubbaRao {et~al.}(2002)SubbaRao, Frieman, Bernardi, Loveday, Nichol,
  Castander, \& Meiksin}]{SubbaRao2002}
SubbaRao, M., Frieman, J., Bernardi, M., Loveday, J., Nichol, B., Castander,
  F., \& Meiksin, A. 2002, Proc. SPIE, 4847, 452

\bibitem[{{Tonry} \& {Davis}(1979)}]{TonryDavis1979}
{Tonry}, J., \& {Davis}, M. 1979, \aj, 84, 1511

\bibitem[{Windley(2005)}]{windley11rest}
Windley, P.~J. 2005, Chapter, 11, 237

\bibitem[{{Yuan} {et~al.}(2015){Yuan}, {Lidman}, {Davis}, {Childress},
  {Abdalla}, {Banerji}, {Buckley-Geer}, {Carnero Rosell}, {Carollo},
  {Castander}, {D'Andrea}, {Diehl}, {Cunha}, {Foley}, {Frieman}, {Glazebrook},
  {Gschwend}, {Hinton}, {Jouvel}, {Kessler}, {Kim}, {King}, {Kuehn},
  {Kuhlmann}, {Lewis}, {Lin}, {Martini}, {McMahon}, {Mould}, {Nichol},
  {Norris}, {O'Neill}, {Ostrovski}, {Papadopoulos}, {Parkinson}, {Reed},
  {Romer}, {Rooney}, {Rozo}, {Rykoff}, {Sako}, {Scalzo}, {Schmidt}, {Scolnic},
  {Seymour}, {Sharp}, {Sobreira}, {Sullivan}, {Thomas}, {Tucker}, {Uddin},
  {Wechsler}, {Wester}, {Wilcox}, {Zhang}, {Abbott}, {Allam}, {Bauer},
  {Benoit-L{\'e}vy}, {Bertin}, {Brooks}, {Burke}, {Carrasco Kind},
  {Covarrubias}, {Crocce}, {da Costa}, {DePoy}, {Desai}, {Doel}, {Eifler},
  {Evrard}, {Fausti Neto}, {Flaugher}, {Fosalba}, {Gaztanaga}, {Gerdes},
  {Gruen}, {Gruendl}, {Honscheid}, {James}, {Kuropatkin}, {Lahav}, {Li},
  {Maia}, {Makler}, {Marshall}, {Miller}, {Miquel}, {Ogando}, {Plazas},
  {Roodman}, {Sanchez}, {Scarpine}, {Schubnell}, {Sevilla-Noarbe}, {Smith},
  {Soares-Santos}, {Suchyta}, {Swanson}, {Tarle}, {Thaler}, \&
  {Walker}}]{fang2015}
{Yuan}, F. {et~al.} 2015, \mnras, 452, 3047

\end{thebibliography}
\bibliographystyle{hapj}

\end{document}